\documentclass[preprint,11pt,authoryear,a4paper]{elsarticle}

\usepackage[utf8]{inputenc}
\usepackage{amssymb}
\usepackage{amsmath}
\usepackage{amsfonts}
\usepackage{array,arydshln}
\usepackage{booktabs} 
\usepackage{multirow}
\usepackage[figuresright]{rotating}
\usepackage{pdflscape}
\usepackage{natbib}
\usepackage{color}
\usepackage{pifont}
\usepackage{float}
\usepackage[hypcap]{caption}
\usepackage{textcomp}
\usepackage{calc}
\usepackage{setspace}
\usepackage{enumitem}
\usepackage{xfrac}
\usepackage{makecell}

\usepackage{booktabs} 
\usepackage{atbegshi} 

\usepackage{float}
\floatstyle{plaintop}
\newfloat{myalgorithm}{!tbp}{lop}
\floatname{myalgorithm}{Algorithm}
\usepackage{ulem}
\usepackage{listings}
\definecolor{colorlink}{rgb}{0, 0, .6}  
\definecolor{darkgreen}{rgb}{0, .5, 0}  
\definecolor{orange}{rgb}{.7, .40, 0}   
\definecolor{grey}{rgb}{.6, .6, .6}     

\newcounter{sidewaystablepage}
\AtBeginShipout{%
  \ifnum\value{sidewaystablepage}>0
    \thispagestyle{empty}%
    \addtocounter{sidewaystablepage}{-1}
  \fi
}

\newcommand{\lspac}{1.20}  

\usepackage[pdftex]{hyperref}
\hypersetup{colorlinks=true,citecolor=colorlink,filecolor=colorlink,linkcolor=colorlink,urlcolor=colorlink}

\journal{ar$\chi$iv}

\begin{document}
\setcounter{page}{1}
\begin{frontmatter}

\title{\textbf{False Discovery Rate and Localizing Power}}

\author[utrgv]{Anderson M.\ Winkler}\ead{anderson.winkler@utrgv.edu}\corref{corresponding}
\author[nih]{Paul A.\ Taylor}\ead{paul.taylor@nih.gov}
\author[big]{\\Thomas E.\ Nichols}\ead{thomas.nichols@bdi.ox.ac.uk}
\author[usc]{Chris Rorden}\ead{rorden@mailbox.sc.edu}
\cortext[corresponding]{Corresponding author.}

\address[utrgv]{Department of Human Genetics,\\The University of Texas Rio Grande Valley, Brownsville, Texas, \textsc{usa}}
\address[nih]{Scientific and Statistical Computing Core, National Institute of Mental Health,\\National Institutes of Health, Bethesda, Maryland, \textsc{usa}.}
\address[big]{Big Data Institute, University of Oxford, Oxford, \textsc{uk}}
\address[usc]{Department of Psychology, University of South Carolina,\\Columbia, South Carolina, \textsc{usa}.}

\begin{abstract}
False discovery rate (FDR) is commonly used for correction for multiple testing in neuroimaging studies. However, when using two-tailed tests, making directional inferences about the results can lead to a vastly inflated error rate, even approaching 100\% in some cases. This happens because FDR controls the error rate only globally, over all tests, not within subsets, such as among those in only one or another direction. Here we consider and evaluate different strategies for FDR control in such cases, using both synthetic and real imaging data. Approaches that separate the tests by direction of the hypothesis test, or by the direction of the resulting test statistic, more properly control the directional error rate and preserve FDR benefits, albeit with a doubled risk of errors under complete absence of signal. Strategies that combine tests in both directions, or that use simple two-tailed p-values, can lead to invalid directional conclusions, even if these tests remain globally valid. A solution to this problem is through the use of selective inference, whereby positive and negative tails are treated as sets (families), which are screened locally, then subjected to FDR at a modified level that controls average FDR over those that survive the initial screening. Moreover, the BKY procedure can be used in place of the well-known Benjamini-Hochberg, yielding additional power. These methods are easy to implement. Finally, to enable valid thresholding for directional inference, we suggest that imaging software should allow the user to set asymmetrical thresholds for the two sides of the statistical map. While FDR continues to be a valid, powerful procedure for multiple testing correction, care is needed when making directional inferences for two-tailed tests, or more broadly, when making any localized inference.
\end{abstract}

\end{frontmatter}

\renewcommand\floatpagefraction{.001}
\makeatletter
\setlength\@fpsep{\textheight}
\makeatother

\vspace{1cm}


\section{Introduction}
\label{sec:intro}
\setstretch{\lspac}

In the context of the multiple testing problem, control over the \textit{false discovery rate} (\textsc{fdr}) provides an alternative to the control over the \textit{familywise error rate} (\textsc{fwer}): while \textsc{fwer} refers to the chance of even one type \textsc{i} (false positive) error is committed in a set (family) of statistical tests, \textsc{fdr} refers to the proportion of such errors among tests for which the null hypothesis has been rejected, that is, the proportion of false discoveries among all discoveries. These topics have been extensively covered in the literature; for an early review of the multiple testing problem in brain imaging, see \citet{Nichols2003}; for a review in genetics, also broadly applicable to imaging, see \citet{Goeman2014}; for early papers and for an overview of the development of \textsc{fdr}, see \citet{Seeger1968, Soric1989, Benjamini2000}; for its introduction to neuroimaging, see \citet{Genovese2002}; for a critique, see \citet{Chumbley2009}; for a discussion of various \textsc{fdr} approaches, see \citet{Korthauer2019}. Methods for \textsc{fwer} correction in brain imaging are often based on the random field theory \citep[\textsc{rft} --][]{Worsley1996, Worsley2004}, permutation tests \citep{Holmes1996, Nichols2002, Winkler2014}, Monte Carlo simulations \citep{Poline1993, Forman1995, Cox1996} and occasionally, the method based on \citet{Bonferroni1936}.

Two hallmarks of most neuroimaging studies are that they can involve thousands of statistical tests, and that one can reasonably expect effects to be observed in both directions for many of these. Controlling the \textsc{fdr} aims to provide more power than \textsc{fwer} in the setting where many null hypotheses are false. This benefit has seen \textsc{fdr} widely adopted in neuroimaging. However, the behavior of \textsc{fdr} when signal manifests in positive direction for some tests, and negative for others, has not received as much attention, even if it is a pertinent problem in the field. In principle, using \textsc{fdr} with two-tailed tests is straightforward: the two-tailed p-values are obtained, then an \textsc{fdr}-controlling procedure is applied. As long as inference remains focused on \textit{all} tests, using \textsc{fdr} on two-tailed p-values has nothing special compared to using one-tailed p-values. Using two-tailed tests can become problematic if, after doing the test, the researcher proceeds to draw inferences on each direction: \textsc{fdr} is not controlled within one direction or another, only globally. Previous literature on directional hypotheses relevant to this problem can be found in \citet{Shaffer1995, Shaffer2002, Williams1999, Benjamini2005}. The same problem emerges if a researcher attempts to draw inferences within regions of a map of p-values after applying \textsc{fdr} globally across all of them. Both problems are a general manifestation of the fact that \textsc{fdr}, while a powerful procedure, does not have the ability to allow localization within subsets of tests, that is, it lacks \textit{localizing power}. As one of our reviewers pointed out, ``making discoveries in either direction while controlling the \textsc{fdr} is trickier than it may seem''.

In this note we start by briefly revisiting the multiple testing problem in imaging, p-value adjustment, the use of \textsc{fdr}, and discuss interpretation for two-tailed tests in the light of statistical properties of \textsc{fdr}-controlling procedures. We demonstrate that merely applying \textsc{fdr} does not control the error rate on each side of the statistical map when the researcher wishes to make inferences about the direction after using two-tailed tests, and can lead to invalid conclusions when effects are asymmetrically distributed across the two sides. We investigate methods that can be considered to address such issues, emphasizing the method proposed by \citet{Benjamini2014} for selective inference. We also bring attention to the \textsc{fdr} approach proposed by \citet{Benjamini2006}, which is more powerful than the one originally proposed by \citet{Benjamini1995}, and which has seldom been used in brain imaging. We also offer a method for p-value adjustment using this more powerful procedure. While the paper is presented in a neuroimaging context, the same topics are relevant to other fields in which \textsc{fdr} is widely used.

\section{Theory}
\label{sec:theory}

\subsection{Preliminaries}

Consider a number $V$ of voxels (or vertices, or edges, or faces) of a representation (image) of the brain. For each voxel, a null hypothesis $\mathcal{H}^{0}_{i}$, $i \in \{i, \ldots, V\}$ is tested through the computation of some test statistic $t_{i}$, and a corresponding (one-tailed) p-value $p_{i}$ that represents the probability of finding $t_{i}$ at least as large as the one observed. Each p-value can be considered statistically significant if equal or below some pre-defined test level $\alpha$ (typically 0.05). If the distribution of the test statistic is symmetric around zero and continuous, two-tailed p-values can be obtained from one-tailed as $p^{\text{two}}=1-\left|2p^{\text{one}}-1\right|$. More generally, if the distribution is either continuous or discrete, two-tailed p-values can be obtained as $p^{\text{two}} = 2\cdot\min\left\{p^{\text{one}},1-p^{\text{one}}+C\right\}$, where $C=0$ for continuous (often parametric) p-values, or $C=1/J$ for discrete, permutation-based (non-parametric) p-values, and $J$ is the number of permutations. If the distribution of the test statistics is asymmetric, two-tailed p-values must be obtained from that distribution, even if one-tailed p-values are available.

\subsection{The multiple testing problem}

Simply comparing each p-value to $\alpha$ leads to many of them to be spuriously considered significant, even if all null hypotheses are true; for example, with $\alpha$ $=$ $0.05$, 5\% of the tests are expected to be found as significant even if no effect is present in any of them. One could address this \textit{multiple testing problem} by choosing a more stringent test level. For example, $\alpha$ could be chosen such that it refers not to errors in each individual test, but to any error in the whole set of tests performed. If all tests are independent, this more stringent level could be computed using the \v{S}id\'ak correction: $\alpha^{\text{\v{S}id\'ak}} = 1 - (1-\alpha)^{1/V}$. Before the computer era, finding the arbitrary root of a number was time-consuming and impractical, particularly for high-degree roots. A simpler method that does not require complex computations, even if slightly more conservative than that of \text{\v{S}id\'ak}, is based on the so-called Bonferroni inequalities \citep{Bonferroni1936, Hochberg1987}, and uses as test level $\alpha^{\text{Bonf.}} = \alpha/V$. The degree of conservativeness of Bonferroni over \text{\v{S}id\'ak} is itself a function of $\alpha$ \citep{Alberton2020}. Bonferroni and \text{\v{S}id\'ak} methods control the chance that \textit{any} test will be found as significant among a set of tests, i.e., they control the \textsc{fwer}. Both methods can be remarkably conservative if the tests are positively dependent, as typical in brain imaging.

Regardless of the degree of dependency among tests, it may be the case that, faced with the task of screening many such tests, a researcher may be willing to tolerate a small amount of them to be incorrectly declared significant (false discoveries), as long as a great majority of tests are correctly declared as such (true discoveries), and the errors are limited; if no discoveries are made, there are no false positives by definition. In other words, the researcher may be willing to accept \textit{some} false discoveries as long as their proportion over all discoveries does not spill over some pre-defined threshold $q$ (often $q$ $=$ $0.05$). Controlling this proportion is the essence of \textsc{fdr} approaches.

\subsection{The basic \textsc{fdr} procedure: \textsc{bh}}
\label{sec:bh}

Control over the \textsc{fdr} can be achieved with the procedure by \citet{Benjamini1995} (\textsc{bh}): let $p_{(1)} \leqslant \ldots \leqslant p_{(V)}$ be the ordered p-values. Find the largest index $i$ for which:

\begin{equation}
p_{(i)} \leqslant \frac{iq}{V}
\label{eqn:bh}
\end{equation}

\noindent
where $q$ is the upper limit on the proportion of false discoveries that the researcher is willing to tolerate. Then reject all null hypotheses that have p-values smaller than or equal to that critical $p_{(i)}$. Graphically, one would lay the ordered p-values in a Cartesian plane that has as horizontal axis the index $i$; draw a line with slope $q/V$ passing through the origin, then reject all tests whose ordered p-values lie to the left of the highest p-value at or below that line (Figure~\ref{fig:thresholds}, left panel).

When the tests are independent, the \textsc{bh} procedure controls the \textsc{fdr} at the level $qV_0/V$, where $V_0$ is the (unknown) number of true null hypotheses. However, complete independence is not strictly necessary: \citet{Benjamini2001} showed that it is sufficient that tests exhibit a form of dependence that they termed \textit{positive regression dependency on subsets} (\textsc{prds}). In this form of dependence, knowledge that one p-value where the null is true is small does not make less likely that the other $(V-1)$ p-values are also small; additional literature on forms of dependency relevant to \textsc{prds} can be found in \citet{Lehmann1966, Esary1967, Sarkar1969, Kimeldorf1989}. Analyses where a false positive does not make the other tests less likely to be found significant, as typical in brain imaging, satisfy the \textsc{prds} condition. If \textsc{prds} is satisfied, the \textsc{bh} procedure still controls the \textsc{fdr}.

A procedure that controls the \textsc{fdr} more generally, for any dependence between tests (as opposed to only under independence or \textsc{prds}) was proposed by \citet{Benjamini2001} (\textsc{by}). The procedure is similar to \textsc{bh}, except for a small modification in Equation~\ref{eqn:bh}, which becomes $p_{(i)} \leqslant iq/V/c(V)$, where $c(V) = \sum^{V}_{j=1}1/j$. Graphically, this results in a less inclined threshold line, and thus, a smaller cutoff $p_{(i)}$ (more conservative) than with the original \textsc{bh} procedure.

\subsection{A more powerful \textsc{fdr} procedure: \textsc{bky}}
\label{sec:bky}

Unfortunately, \textsc{bh} and \textsc{by} are both conservative in the sense that the expected \textsc{fdr} is below $qV_0/V$, which is always less than or equal to the desired level $q$. To mitigate this conservativeness, Benjamini, Krieger, and Yekutieli proposed to estimate $V_0$ after a first pass using $\textsc{bh}$ with $q' = q/(1+q)$, then repeat with $q^{*}=q' V/\hat{V}_0$ if at least one hypothesis was rejected in the first pass \citep[or simply \textsc{bky}]{Benjamini2006}. This two-pass procedure has been proposed for use in brain imaging \citep{Chen2009}, albeit it remains rarely used. In their original paper, \citet{Benjamini2006} recognised that there is no reason to stop after the second pass; in fact, the procedure can be repeated iteratively until there are no more rejections. Doing so leads to following procedure (\textsc{bky}), which we adopt here: starting from the lowest p-value, keep rejecting hypotheses for the successive $p_{(i)}$ as long as at least one $p_{(j)}$, $j \geqslant i$, satisfies:

\begin{equation}
p_{(j)} \leqslant \frac{jq}{V+1-i(1-q)}
\label{eqn:bky}
\end{equation}

\noindent
stopping at the first $i$ for which the inequality is no longer satisfied for any $j \geqslant i$; all subsequent nulls, inclusive, are retained \citep[Definition 7]{Benjamini2006}.

\begin{figure}[!p]
\centering
\hspace*{-0.25\textwidth}\includegraphics[width=1.5\textwidth]{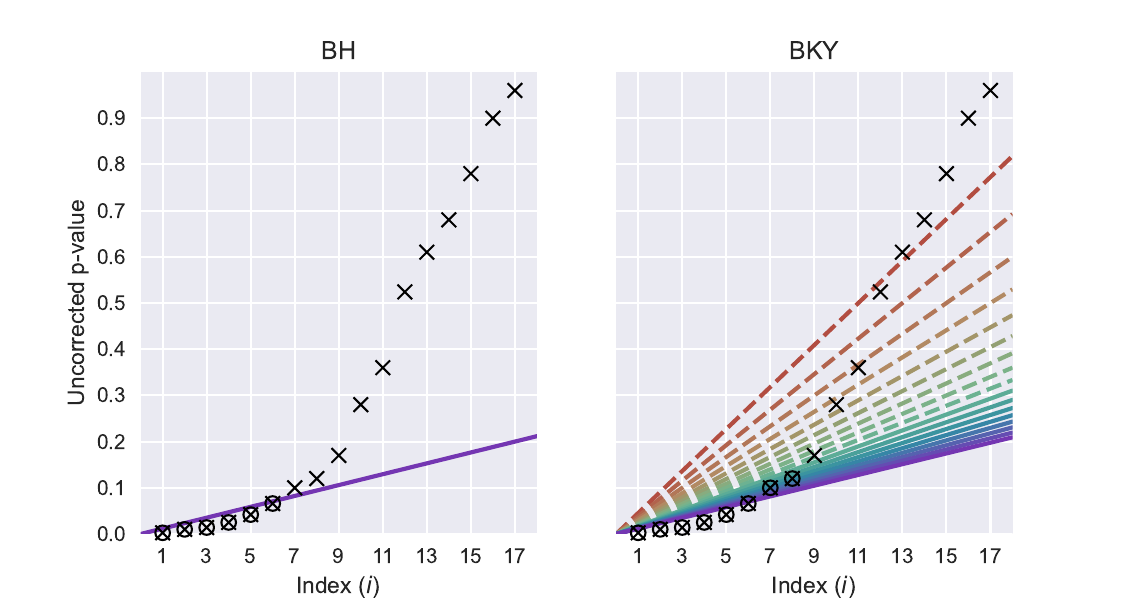}
\caption{Comparison of two \textsc{fdr} methods, \citet{Benjamini1995} (\textsc{bh}) and \citet{Benjamini2006} (\textsc{bky}), for $q=0.20$. Each symbol $\boldsymbol{\times}$ represents an uncorrected p-value; these are placed in ascending order ($i$); the p-values considered significant according to the respective method are surrounded by a circle $\boldsymbol{\otimes}$. In \textsc{bh}, p-values smaller than or equal to the highest p-value below a single inclined threshold (blue line) are declared significant. In \textsc{bky}, the thresholds themselves are adaptive: their inclination increases for larger p-values (colored lines), making it easier to detect significant p-values, provided that smaller ones are also found significant (for clarity, we show in dashed lines the thresholds for tests considered not significant). The p-values used in this example are: \{0.0026, 0.01, 0.014, 0.025, 0.042, 0.066, 0.1, 0.12, 0.17, 0.28, 0.36, 0.524, 0.61, 0.68, 0.78, 0.9, 0.96\}.}
\label{fig:thresholds}
\end{figure}

As with \textsc{bh}, \textsc{bky} can also be interpreted graphically (Figure~\ref{fig:thresholds}, right panel): for each ordered p-value $p_{(i)}$, trace a threshold line with slope $q/(V+1-i(1-q))$ passing through the origin, and check if any p-value from that position (inclusive) onward is at or below that line; if so, reject the corresponding null hypothesis, and move to the next p-value, now with a new threshold line that is steeper (more liberal) than the previous. The \textsc{bky} procedure differs from \textsc{bh}, however, not only on the use of an adaptive threshold line, whose slope varies according to the position $i$, but also on the stopping criterion: whereas \textsc{bh} uses as cutoff the largest $p_{(i)}$ that satisfies its inequality, \textsc{bky} stops at the first $p_{(i)}$ that no longer satisfies it. This has importance when computing adjusted p-values, discussed next.

\subsection{Adjustment of p-values}
\label{sec:adjustment}

It is convenient to inspect a map of p-values that incorporates correction for multiple testing: thresholding such map at the desired test level immediately reveals significant results and is straightforward to examine in most imaging software via, e.g., a slider, or by varying the upper or lower limits of the color scale, or by changing image contrast window center and width. In contradistinction, maps of uncorrected p-values need be thresholded at some corrected $\alpha$ or $q$ levels. Corrected p-values for the \textsc{bh} procedure can be computed by finding the smallest $q$ over every position $i$ that still satisfies the inequality given in Equation~\ref{eqn:bh}:

\begin{equation}
q_{(i)}^\textsc{bh} \geqslant \frac{p_{(i)}V}{i} = \left[p_{(i)}^\textsc{bh}\right]_{\text{corr}}
\label{eqn:bh_corr}
\end{equation}

\noindent
where the inequality becomes equality at the lowest (strictest) value of $q_{(i)}$ for which the respective $p_{(i)}$ is significant (the notation with brackets $[$~$]$ avoids double subscripts). In a similar manner, for \textsc{bky}, corrected p-values can be found by solving Equation~\ref{eqn:bky} for $q$ at $i$:

\begin{equation}
q_{(i)}^\textsc{bky} \geqslant \min_{j \geqslant i}\left\{\frac{p_{(i)}(V+1-i)}{j-ip_{(i)}}\right\} = \left[p_{(i)}^\textsc{bky}\right]_{\text{corr}}
\label{eqn:bky_corr}
\end{equation}

Unfortunately, corrected p-values for \textsc{bh} and \textsc{bky} are not guaranteed to be monotonically related to their corresponding uncorrected p-values (that is, corrected p-values not necessarily increase or decrease as their corresponding uncorrected p-values increase or decrease). Graphically, the corrected p-values correspond to the slope of the line that connects each point to the center of the Cartesian grid, and that slope can be higher than $q$ even for p-values considered significant. Such lack of monotonicity is problematic because thresholding a map of such corrected p-values at some desired level $q$ will not produce the same result as thresholding the map of original p-values at the critical $p_{(i)}$ identified with the respective \textsc{fdr}-controlling method. One may expect monotonicity when doing statistical tests: it allows one to trade-off between maximizing power while minimizing false positives \citep{Zeisel2011}. Following \citet[Definition 2.4]{Yekutieli1999}, \textit{adjusted} p-values, which replace corrected p-values without having this issue, can be computed for the \textsc{bh} procedure as:

\begin{equation}
\left[p_{(i)}^\textsc{bh}\right]_{\text{adj}} = \min_{j \geqslant i}\left\{ \left[p_{(j)}^\textsc{bh}\right]_{\text{corr}} \right\}
\label{eqn:bh_adj}
\end{equation}

\noindent
which can be read as the cumulative minimum of the corrected p-values, starting from the largest. Because \textsc{bky} uses a different stopping criterion than \textsc{bh}, the adjustment is also different: rather than using the smallest corrected p-value from the current position $i$ onward, \textsc{bky} requires the largest corrected p-value up to the current position $i$, that is, the cumulative maximum:

\begin{equation}
\left[p_{(i)}^\textsc{bky}\right]_{\text{adj}} = \max_{j \leqslant i}\left\{ \left[p_{(j)}^\textsc{bky}\right]_{\text{corr}} \right\}
\label{eqn:bky_adj}
\end{equation}

\begin{figure}[!p]
\centering
\hspace*{-0.25\textwidth}\includegraphics[width=1.5\textwidth]{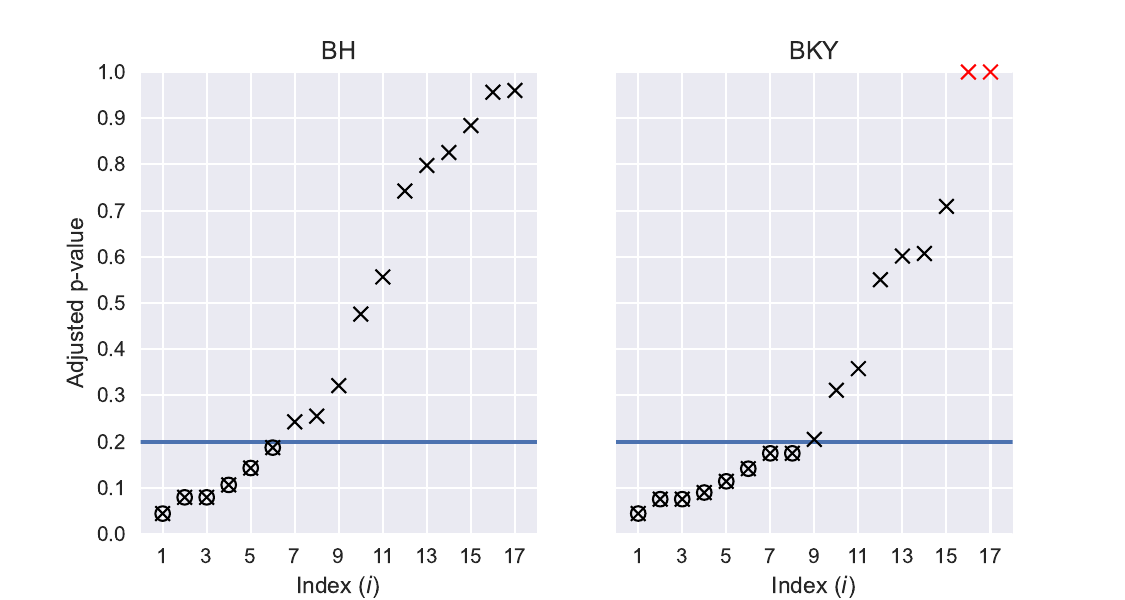}
\caption{Comparison of \textsc{fdr}-adjusted p-values produced with  \citet{Benjamini1995} (\textsc{bh}) and \citet{Benjamini2006} (\textsc{bky}) methods, for the same data shown in Figure~\ref{fig:thresholds}. Each symbol $\boldsymbol{\times}$ represents a uncorrected p-value; these are placed in ascending order ($i$); the p-values considered significant are surrounded by a circle $\boldsymbol{\otimes}$. The adjustment ensures that a single, fixed threshold can be applied to all tests (blue line; $q=0.20$); values below that threshold are considered significant. Note that adjusted p-values can sometimes be greater than 1; these can be replaced by 1 (markers \textcolor{red}{$\boldsymbol{\times}$} in red).}
\label{fig:adjusted}
\end{figure}

With adjusted p-values, significance thresholds no longer depend on the position $i$ for \textsc{bky}: maps of adjusted p-values can be arbitrarily thresholded at any $q$ in the interval between 0 and 1 (Figure~\ref{fig:adjusted}). It should be noted that ties are very common with adjusted p-values; they pose no problem for interpretation. Interestingly, using Equation~\ref{eqn:bky_adj} for \textsc{bky}, as opposed to a formulation equivalent to Equation~\ref{eqn:bh_adj}, solves the issue with the paradoxical results observed by \citet{Reiss2012}, whereby adjusted p-values would seem smaller than their unadjusted counterparts.

For \v{S}id\'ak, Bonferroni, and permutation methods, \textsc{fwer}-corrected p-values are monotonically related to their underlying p-values, such that no distinction between correction or adjustment needs to be made. \text{\v{S}id\'ak}-adjusted p-values can be computed as $\left[p_i^{\text{\v{S}id\'ak}}\right]_{\text{adj}} = 1 - (1-p_i)^V$, whereas Bonferroni-adjusted p-values can be computed as $\left[p_i^{\text{Bonf.}}\right]_{\text{adj}} = p_iV$. For permutation methods, adjusted p-values are computed by referring to the distribution of the maximum statistic across tests \citep{Westfall1993, Winkler2014}. Note that adjusted p-values computed for various \textsc{fdr} procedures or for Bonferroni can be greater than 1; this can be treated by replacing adjusted p-values greater than 1 by 1; in Figure~\ref{fig:adjusted} these are marked in red.

\subsection{Localizing power}
\label{sec:theory_localizing}

Methods for \textsc{fwer} and \textsc{fdr} control differ not only on the measurement of error that they target, but on at least another aspect: if a procedure controls the \textsc{fdr} at some level for a set of hypotheses, this control does not necessarily extend to every subset of hypotheses, and the actual \textsc{fdr} can in fact be much higher (or even much lower) for any subset; a procedure that controls the \textsc{fwer} for a set of hypotheses, however, also controls the \textsc{fwer} for every subset. This feature, that \textsc{fwer} procedures have, but \textsc{fdr} procedures have not, was described by \citet{Finner2001}, and called \textit{subsetting property} by \citet{Goeman2014}. In the context of imaging, we find the term \textit{localizing power} (or the lack of it) preferable as it immediately conveys the meaning that further localization within parts of an image may or may not be possible, depending on the procedures used for inference \citep{Friston1996, Nichols2003, Noble2022}.

The ability to localize effects within parts of an image requires that the error rate is controlled within that part. Not having localizing power is problematic, for example, if the researcher wishes to draw conclusions within specific regions after \textsc{fdr} has been applied globally. Likewise, this becomes problematic if the researcher wishes to draw conclusions pertaining to the direction of effects --- that is, inferring about each tail separately --- after applying \textsc{fdr} on two-tailed p-values. To illustrate that, consider these two hypothetical examples:

\paragraph{Example 1: Direction error} An investigation into the association between cortical thickness (measured at every vertex of a surface representation of the brain) and a laboratory measure involves a sample of hundreds of individuals. Acknowledging that effects could be manifest in either direction, and pursuing statistical rigor, the researcher employs two-tailed p-values, with adjustments for multiple testing using an \textsc{fdr}. The map of test statistics is then masked at the significant level $q=0.01$, thus retaining for visualization only the significant vertices. The analysis uncovers large areas in the frontal and parietal lobes with a positive relationship to the laboratory measure, as hypothesized. However, it also reveals a negative association spanning broad portions of the fusiform gyrus, as well as vertices showing positive or negative associations scattered across the cortex. The researcher conducting the analysis writes a report detailing the novel findings in the negative direction, emphasizing the stringent methodologies that incorporated two-tailed tests and a strict \textsc{fdr} significance threshold, well below the conventional level $q=0.05$, and concludes (erroneously) that only about 1\% of the negative associations are expected to be false positives.

\paragraph{Example 2: Localization error} In a study of a novel compound for reducing the severity of relapses in multiple sclerosis (\textsc{ms}), white matter hyperintensities (\textsc{wmh}) are analyzed voxelwise. Subjects are randomly allocated into active compound and placebo groups. Based on prior knowledge and on random population sampling, normality is assumed and a two-sided group comparison is conducted using a two-sample $t$-test. This produces two-tailed p-values calculated from the Student's $t$-distribution. Control for multiple testing across voxels is achieved using \textsc{fdr} at $q=0.05$. After determining the \textsc{fdr} threshold, the inverse cumulative distribution function of the Student's $t$-distribution is used to derive the critical threshold $t_{\textsc{fdr}}$. Given the two-tailed nature of the test, symmetrical thresholds $+t_{\textsc{fdr}}$ and $-t_{\textsc{fdr}}$ are set in an image visualization tool. The results show a notable reduction in the periventricular \textsc{wmh} load in the group that used the active compound, as expected (and hoped), as well as mixed results in various juxtacortical regions, albeit with mostly increases in  \textsc{wmh} load in these areas. In the report, the investigator concludes (erroneously) that the drug may amplify the \textsc{wmh} load in juxtacortical areas given that, as \textsc{fdr} was used, no more than 5\% of the observed changes in those regions are false positives.

At first, it may seem that the researchers in both examples did everything correctly, going as far as using two-tailed tests and ensuring that the multiplicity of tests would be addressed rigorously. A casual reader of the respective study reports may find no issue. Yet, where it comes to statements about the direction or localization of effects, the interpretation is not that, among the findings in the negative direction (Example~1) or in a particular region (Example~2), the proportion of false positives is within $q$, but only that the proportion of false positives among \textit{all} tests that were subjected to \textsc{fdr} is within $q$. The actual, realized (but unknown) \textsc{fdr} is not guaranteed to be within 1\% for the negative (or positive) direction in Example~1, nor 5\% within the juxtacortical white matter (or any other region) in Example~2, and can in both cases be anywhere between 0 and 100\% in the reported direction or region.

These issues are a consequence of the lack of the subsetting property. Localization errors can be avoided by refraining from making statements or taking actions that could refer to the amount of errors in specific regions when \textsc{fdr} is controlled only globally. The same holds to avoid direction errors. There is, however, a rigorous way to perform localized inferences, and which allows such statements to be made or such actions to be taken. That involves yet another \textsc{fdr}-related procedure, the one proposed by \citet{Benjamini2014}, presented next.

\subsection{Selective inference: \textsc{bb}}
\label{sec:theory_selective}

The problem of directional inference can be framed as a special case of selective inference on multiple sets of hypotheses. First, consider the $V$ tests partitioned into multiple non-overlapping sets (e.g., all voxels within multiple regions of interest, or all tests partitioned by positive or negative test statistics). A researcher could try to bypass the above issues by applying \textsc{fdr} within each set separately \citep[see][for a discussion]{Efron2008}. In the limiting case in which each set is as small as just one voxel, there is nothing left to be corrected within that set, yet the multiple testing problem would still persist through the existence of multiple sets. The researcher could also try to apply \textsc{fdr} within only select, ``promising'' sets of tests, such as those that contain small p-values according to some criteria. This, however, would ignore the selection procedure and lead to inflated error rates in those sets that were selected, unless additional adjustments via, e.g., permutations, were used \citep[see][for one such procedure]{Heller2009}.

To address this, \citet{Benjamini2014} propose to control neither the error rate within each set, nor the global error rate; instead, the problem is addressed by controlling the \textit{average} error rate across the sets. To that goal, they develop the following procedure (\textsc{bb}):

\begin{enumerate}[label=\arabic*)]
\item Compute uncorrected p-values $p_i$ for each test.
\item Define the $S$ non-overlapping sets of tests.
\item Apply a screening procedure within each set. Any procedure that depends only on the p-values within that set can be considered, whether controlling \textsc{fdr} or \textsc{fwer}. Procedures such as \textsc{bh} or, equivalently, \citet{Simes1986}, are fast to compute and are acceptable choices. This is the ``first stage'' procedure.
\item Count the number $R$ of sets in which at least one test was rejected in the screening procedure.
\item If $R=0$, declare no discoveries and stop. Otherwise, apply an \textsc{fdr}-controlling procedure, such as \textsc{bh} or \textsc{bky}, within each selected set using a modified level $q' = qR/S$ to control the expected average \textsc{fdr} over the selected sets. This is the ``second stage'' procedure.
\end{enumerate}

\noindent
\citet{Benjamini2014} prove that if the selection of a set (in the paper termed \textit{family}) depends only on the data within that set, then applying an \textsc{fdr}-controlling procedure at level $q' = qR/S$ within each selected set guarantees that the average false discovery rate across sets is controlled at $q$. The only requirement is that the screening procedure is local within each set, that is, selection is not a function of p-values from the other sets. Hence, for the Example 1 from the previous section, the problem can be solved by partitioning the statistical map into two disjoint sets, one with the positive test statistics, the other with the negative test statistics, then applying a selection procedure (which can be an \textsc{fdr} procedure) to each side of the map; if nothing survives in either side, the researcher stops. If there is at least one rejection in one ($R=1$) or in each ($R=2$) side, proceed to an \textsc{fdr} procedure in each side, this time using level $q'=qR/S$, where $S=2$ given the researcher partitioned the $V$ voxels into two sets. For the Example 2, the problem can be solved by partitioning the map into at least two regions of interest (i.e., two sets, these comprising periventricular and juxtacortical white matter voxels), then if interest is also on making inferences separately across the map, further split each region into positive and negative sides after the test statistics have been computed for each voxel, then subject all such sets of voxels to the \textsc{bb} procedure. In either example, partitioning by the sign of the test statistic works because the sign is independent of the two-tailed p-values; the latter are distributed uniformly between 0 and 1 and, in either side of the map, more extreme test statistics (stronger effects) imply lower p-values. Thus, the requirements for \textsc{fdr}, using either \textsc{bh} or \textsc{bky}, are satisfied. The \textsc{fdr} can then be controlled at the desired level $q$ for the positive and for the negative sides of the map, and further, as in Example 2, for the multiple regions in consideration. Note that, with \textsc{bb}, adjusted p-values can still be computed as described in Section~\ref{sec:adjustment}; only the threshold needs to be modified as $q' = qR/S$.


\section{Evaluation}
\label{sec:evaluation}

\subsection{Synthetic data}

To assess the realized (empirical) \textsc{fdr} across procedures we used synthetic data formed by 2000 tests, which could represent voxels in a group-level result image, or any other large set of tests. For each test, a statistic $z$ was simulated following a normal distribution with zero mean and unit variance. Five different scenarios were considered for independent test statistics: (\textsc{i}) complete null, with no synthetic effects added; (\textsc{ii}) positive effects only, in which 25\% of the tests had their $z$ statistic increased by 3; (\textsc{iii}) negative effects only, in which 25\% of the tests had their $z$ statistic decreased by 3; (\textsc{iv}) both positive and negative effects, balanced, in which 25\% of the tests had their $z$ statistic increased by 3, another 25\% had their $z$ statistic decreased by 3; and (\textsc{v}) both positive and negative effects, unbalanced, in which 10\% of the tests had their $z$ statistic increased by 3, another 40\% had their $z$ statistic decreased by 3. Another five scenarios (\textsc{vi}--\textsc{x}) were considered; these were identical to the first five except that instead of independent tests, the 2000 test statistics were made non-independent, with a correlation of 0.25. In each scenario, uncorrected p-values were computed by referring to the normal cumulative distribution function. A summary of the scenarios is shown in Table~\ref{tab:scenarios}.

\begin{table}[!tbp]
\caption{Scenarios used to evaluate the different \textsc{fdr} approaches with synthetic data. In each scenario, 2000 test statistics were simulated and subjected to \textsc{fdr} procedures. The process was repeated 2000 times. Scenarios \textsc{i}--\textsc{v} contain independent tests; tests in scenarios \textsc{vi}--\textsc{x} have nonzero dependency, as indicated by the ``Correlation'' column.}
\begin{center}
{\small
\hspace*{-0mm}\begin{tabular}{@{}cccc@{}}
\toprule
\shortstack{Simulation\\scenario} & \shortstack{Tests with\\positive effects} & \shortstack{Tests with\\negative effects} & \shortstack{Correlation\\among tests}\\ 
\midrule
\textsc{i}            & 0\%   & 0\%   & 0\\
\textsc{ii}           & 25\%  & 0\%   & 0\\
\textsc{iii}          & 0\%   & 25\%  & 0\\
\textsc{iv}           & 25\%  & 25\%  & 0\\
\textsc{v}            & 10\%  & 40\%  & 0\\
\textsc{vi}           & 0\%   & 0\%   & 0.25\\
\textsc{vii}          & 25\%  & 0\%   & 0.25\\
\textsc{viii}         & 0\%   & 25\%  & 0.25\\
\textsc{ix}           & 25\%  & 25\%  & 0.25\\
\textsc{x}            & 10\%  & 40\%  & 0.25\\
\bottomrule
\end{tabular}}
\end{center}
\label{tab:scenarios}
\end{table}

Simulations for these 10 scenarios were repeated (realized) 2000 times. Each was assessed with both \textsc{bh} and \textsc{bky}. Six different strategies were considered, described below and summarized in Table~\ref{tab:strategies}. Hence, we have 120 sets of results (10 scenarios, 2 \textsc{fdr} procedures, and 6 strategies), each based on the 2000 realizations of random test statistics. For this simulated evaluation, we use $q=0.05$.

\begin{itemize}
\item \textit{Canonical}, in which adjustment is applied to one-tailed p-values in one direction, and to one-tailed p-values in the opposite direction, that is, applied separately to $p$ and to $1-p$. This procedure produces two complete maps of adjusted p-values, one for each direction; alternatively, it produces two thresholds, one for each direction.
\item \textit{Combined}, in which adjustment is applied simultaneously to one-tailed p-values in the two directions, i.e., on the set formed by $p$ and $1-p$. This procedure produces two complete maps of adjusted p-values, one for each direction; alternatively, it produces one threshold, to be applied to both directions.
\item \textit{Two-tailed}, in which two-tailed p-values are subjected to adjustment. This procedure produces one complete map of adjusted p-values; alternatively, it produces one threshold.
\item \textit{Split-tails}, in which the tests are separated into two disjoint sets, one with positive ($z>0$) and another with negative ($z<0$) statistics; then the two-tailed p-values within each set are subjected to adjustment. For the cases in which the test statistic is zero ($z=0$), the uncorrected two-tailed p-value is 1, and so is the adjusted p-value; such \textit{pin risk} is extremely unlikely with continuous test statistics. This procedure produces one complete map of adjusted p-values after the two sets with adjusted p-valuer are joined; alternatively, it produces two thresholds, one for positive test statistics, one for negative test statistics.
\item \textit{Canonical + \textsc{bb}}, in which adjustment is applied to one-tailed p-values in one direction, and to one-tailed p-values in the opposite direction, similarly to the canonical strategy, but using the \textsc{bb} procedure, that is, one set is formed by $p$ and the other by $1-p$. This procedure produces two complete maps of adjusted p-values, one for each direction; alternatively, it produces two thresholds, one for each direction.
\item \textit{Split-tails + \textsc{bb}}, in which the tests are separated into two disjoint sets, one with positive ($z>0$) and another with negative ($z<0$) statistics, similarly as the split-tails strategy. Then the two-tailed p-values within each subset are subjected to adjustment following the \textsc{bb} procedure. This strategy produces one complete map of adjusted p-values after the two sets with adjusted p-values are joined; alternatively, it produces two thresholds, one for positive test statistics, one for negative test statistics. The same considerations regarding pin risk as in the Split-tails method hold. 
\end{itemize}

For the two strategies that used \textsc{bb}, the first stage used the Simes procedure with test level 0.05. Note that canonical and split-tails can naturally be subjected to \textsc{bb} as these two strategies inherently involve two sets of tests; the same does not apply to combined or two-tailed.

\begin{table}[!tbp]
\caption{Summary of the six strategies considered for running \textsc{fdr}.}
\begin{center}
{\small
\hspace*{-0mm}\begin{tabular}{@{}
m{28mm}<{\raggedright}@{\hspace{1mm}}
m{50mm}<{\raggedright}@{\hspace{1mm}}
m{45mm}<{\raggedright}@{\hspace{1mm}}
@{}}
\toprule
\shortstack{Strategy}     & \shortstack{Inputs to \textsc{fdr}}   & \shortstack{Calls to \textsc{fdr}}\\ 
\midrule
Canonical                 & $\{p^{\text{one}}_{i}\}$, $\{1-p^{\text{one}}_{i}\}$  & Two, each with $V$ tests\\
Combined                  & $\{p^{\text{one}}_{i}, 1-p^{\text{one}}_{i}\}$        & One, with $2V$ tests\\
Two-tailed                & $\{p^{\text{two}}_{i}\}$                              & One, with $V$ tests\\
Split-tails               & $\{p^{\text{two}}_{i}: z_{i}>0\}$, $\{p^{\text{two}}_{i}: z_{i}<0\}$  & Two, with a total of $V$ tests\\
Canonical + \textsc{bb}   & $\{p^{\text{one}}_{i}\}$, $\{1-p^{\text{one}}_{i}\}$  & Two*, each with $V$ tests\\
Split-tails + \textsc{bb} & $\{p^{\text{two}}_{i}: z_{i}>0\}$, $\{p^{\text{two}}_{i}: z_{i}<0\}$  & Two*, with a total of $V$ tests\\
\bottomrule
\end{tabular}}
\end{center}
{\footnotesize
For each synthetic voxel $i \in \{1, 2, \ldots, V\}$, $p^{\text{one}}_{i}$ refers to their one-tailed p-values, $p^{\text{two}}_{i}$ refers to their two-tailed p-values; $z_{i}$ are their corresponding test statistics (for example, it could be the result of a signed statistical test, such as the $t$-test).\newline
* The two calls to \textsc{fdr} refer to one per set (family) in the second stage of the \textsc{bb} procedure; the first stage uses as many calls to \textsc{fdr} or to any other screening procedure as there are sets.\par}
\label{tab:strategies}
\end{table}

\paragraph{Counting true and false positives} For each realization, true and false positives are considered in relation to the true signal; for directional inferences (positive or negative), these refer to the results in the positive and negative sides of the statistical map; for non-directional (both sides), the directions of the test statistic and of the true signal are ignored; these definitions are formalized in Table~\ref{tab:counting}. The proportion of false discoveries within a realization is the number of false discoveries divided by the total number of discoveries.

\paragraph{Assessing power} The power of each of the \textsc{fdr} methods and the six different strategies is the ratio between the number of true discoveries by the number of tests in which true signal was added, computed for reach realization. For directional inferences (positive or negative), these refer to the results in the positive and negative sides of the statistical map; for non-directional (both sides), the directions of the test statistic and of the true signal are ignored; as described in Table~\ref{tab:counting}. Thus, power here represents a distinct quantity in relation to the usual statistical power. Here it represents the sensitivity of a multiple testing procedure judged by its ability to recover a number of tests that contain true signal, that is, tests in which the null hypothesis is false.

\begin{table}[!tbp]
\caption{Counting of tests that meet criteria for containing a true synthetic effect in a given direction, for having the sign of the test statistic matching the direction of the signal, and being statistically significant after an \textsc{fdr}-controlling procedure. These counts are used to calculate the empirical proportion of false discoveries and empirical power with synthetic data.}
\begin{center}
{\small
\begin{tabular}{@{}
m{50mm}<{\raggedright}@{\hspace{1mm}}
m{74mm}<{\raggedright}@{\hspace{1mm}}
@{}}
\toprule
Direction & Counts \\ 
\midrule
\multicolumn{2}{l}{\textit{Both sides:}}\\
True effects      & $\#\{i: \top^{+}_{i} \text{ or } \top^{-}_{i}\}$\\
True discoveries  & $\#\{i: (p^{\text{adj}}_{i} \leqslant q) \text{ and } (\top^{+}_{i} \text{ or } \top^{-}_{i})\}$  \\
False discoveries & $\#\{i: (p^{\text{adj}}_{i} \leqslant q) \text{ and } \text{not }(\top^{+}_{i} \text{ or } \top^{-}_{i})\}$ \\
\midrule
\multicolumn{2}{l}{\textit{Positive side:}}\\
True effects      & $\#\{i: \top^{+}_{i}\}|$\\
True discoveries  & $\#\{i: (p^{\text{adj}}_{i} \leqslant q) \text{ and } (z_{i}>0) \text{ and } \top^{+}_{i}\}$ \\
False discoveries & $\#\{i: (p^{\text{adj}}_{i} \leqslant q) \text{ and } (z_{i}>0) \text{ and } \text{not }\top^{+}_{i}\}$ \\
\midrule
\multicolumn{2}{l}{\textit{Negative side:}}\\
True effects      & $\#\{i: \top^{-}_{i}\}|$\\
True discoveries  & $\#\{i: (p^{\text{adj}}_{i} \leqslant q) \text{ and } (z_{i}<0) \text{ and } \top^{-}_{i}\}$ \\
False discoveries & $\#\{i: (p^{\text{adj}}_{i} \leqslant q) \text{ and } (z_{i}<0) \text{ and } \text{not }\top^{-}_{i}\}$ \\
\bottomrule
\end{tabular}}
\end{center}
{\footnotesize
For each synthetic voxel $i \in \{1, 2, \ldots, V\}$, $\top^{+}_{i}$ and $\top^{-}_{i}$ indicate whether a true positive or true negative effect was added to that voxel, $z_i$ represents the corresponding test statistic at that voxel, and $p^{\text{adj}}_{i} \leqslant q$ represents a significant adjusted p-value at that voxel after an \textsc{fdr} procedure. For each realization, the counts are the sizes (cardinalities) of each of the sets. The empirical \textsc{fdr} within a realization is the number of false discoveries observed in a given direction divided by the total number of discoveries in that direction, whereas power is the number of true discoveries in a given direction divided by the number of tests with true effects in that direction. For computations that refer to both sides, the direction is ignored. Averaging the proportion of false discoveries or the calculated power across realizations yields the results shown in Tables \ref{tab:synthetic_fdp} and \ref{tab:synthetic_pwr}, respectively.\par}
\label{tab:counting}
\end{table}

The overall empirical \textsc{fdr} (observed false discovery proportion) and power were computed by averaging across realizations. Confidence intervals (95\%) were calculated with the Wald method \citep{Brown2001}.

\subsection{Real data}

To assess these six different strategies for two-tailed \textsc{fdr} with real data, as well as to provide an empirical comparison between \textsc{bh} and \textsc{bky}, we use a statistical map originally produced as part of the Neuroimaging Analysis Replication and Prediction Study \citep[\textsc{narps} ---][]{Botvinik-Nezer2020}. The \textsc{narps} was an effort to assess the variability of results among different analysis teams analyzing the same imaging dataset. In the study, the teams were provided with the same raw functional magnetic resonance imaging (f\textsc{mri}) data, collected during a gambling task. Each team analyzed the data to investigate predefined hypotheses about brain activation related to gains and losses, and employed diverse analysis workflows, with no two teams choosing identical pipelines. We use (for no particular reason) the maps produced by team 9U7M for \textsc{narps} Hypothesis 1. Note that the specifics of the hypothesis are not relevant to the present analysis; consult \citet{Botvinik-Nezer2020} for details about data acquisition, the predefined hypotheses, the overall \textsc{narps} methodology, and its data provenance. For this real data evaluation, we use $q=0.05$.


\section{Results}
\label{sec:results}

\subsection{Synthetic data}

Tables \ref{tab:synthetic_fdp} and~\ref{tab:synthetic_pwr} summarize the results for the analyses using synthetic data across the 10 different scenarios that involved varying proportions of true positive and negative effects, as well as dependencies among tests.  For a given test, the ideal result is that the confidence interval includes the chosen $q=0.05$ value (represented in the Table in percentage, i.e., 5\%).

\begin{table}[!b]
\centering
\caption{\textit{(page \pageref{tab:synthetic_fdp_noref})} Comparison of \textbf{error rate} (false discovery proportion, in \%) among different \textsc{fdr} methods and correction strategies, when false positives are counted considering one side (positive or negative) or both sides of the statistical map. These results are based on synthetic data. The nominal proportion of false discoveries is $q\cdot 100\%=5\%$; their corresponding confidence intervals (95\%) are shown between parentheses.}
\label{tab:synthetic_fdp}
\end{table}

\begin{table}[!b]
\centering
\caption{\textit{(page \pageref{tab:synthetic_pwr_noref})} Comparison of \textbf{power} (\%) among different \textsc{fdr} methods and correction strategies, when true positives are counted considering one side (positive or negative) or both sides of the statistical map. These results are based on synthetic data. The nominal proportion of false discoveries is $q\cdot 100\%=5\%$; their corresponding confidence intervals (95\%) are shown between parentheses. For scenarios that did not include signal in the direction shown, there is no power to be calculated; these are marked with a ``---''.}
\label{tab:synthetic_pwr}
\end{table}

\begin{sidewaystable}[!p]
\setcounter{sidewaystablepage}{2}
\begin{center}
{\scriptsize
\hspace*{-24mm}
\begin{tabular}{@{}
p{10mm}@{\hspace{1mm}}
p{20mm}@{\hspace{1mm}}
p{20mm}@{\hspace{1mm}}
p{20mm}@{\hspace{1mm}}
p{20mm}@{\hspace{1mm}}
p{20mm}@{\hspace{1mm}}
p{20mm}@{\hspace{1mm}}
p{20mm}@{\hspace{1mm}}
p{20mm}@{\hspace{1mm}}
p{20mm}@{\hspace{1mm}}
p{20mm}@{\hspace{1mm}}
p{20mm}@{\hspace{1mm}}
p{20mm}@{}}
\toprule
{} & 
\multicolumn{6}{c}{Benjamini--Hochberg} & 
\multicolumn{6}{c}{Benjamini--Krieger--Yekutieli} \\
\cmidrule(r){2-7}\cmidrule(){8-13}
Scn. & 
Canonical & Combined & Two-tailed & Split-tails & Canonical+\textsc{bb} & Split-tails+\textsc{bb} &
Canonical & Combined & Two-tailed & Split-tails & Canonical+\textsc{bb} & Split-tails+\textsc{bb} \\
\midrule
\multicolumn{9}{l}{\textit{Both sides:}}\\
\textsc{i} & 9.1 \scalebox{.7}[1.0]{(7.8--10.4)} & 5.2 \scalebox{.7}[1.0]{(4.3--6.2)} & 5.2 \scalebox{.7}[1.0]{(4.3--6.2)} & 9.1 \scalebox{.7}[1.0]{(7.8--10.4)} & 5.2 \scalebox{.7}[1.0]{(4.3--6.2)} & 5.2 \scalebox{.7}[1.0]{(4.3--6.2)} & 9.3 \scalebox{.7}[1.0]{(8.1--10.6)} & 5.0 \scalebox{.7}[1.0]{(4.0--5.9)} & 5.0 \scalebox{.7}[1.0]{(4.0--5.9)} & 9.3 \scalebox{.7}[1.0]{(8.0--10.6)} & 4.8 \scalebox{.7}[1.0]{(3.9--5.7)} & 4.9 \scalebox{.7}[1.0]{(4.0--5.8)} \\
\textsc{ii} & 3.8 \scalebox{.7}[1.0]{(3.7--3.8)} & 3.8 \scalebox{.7}[1.0]{(3.7--3.8)} & 3.8 \scalebox{.7}[1.0]{(3.7--3.8)} & 3.0 \scalebox{.7}[1.0]{(3.0--3.1)} & 2.0 \scalebox{.7}[1.0]{(1.9--2.0)} & 1.6 \scalebox{.7}[1.0]{(1.6--1.6)} & 4.7 \scalebox{.7}[1.0]{(4.6--4.7)} & 4.1 \scalebox{.7}[1.0]{(4.0--4.1)} & 4.5 \scalebox{.7}[1.0]{(4.4--4.5)} & 4.3 \scalebox{.7}[1.0]{(4.3--4.4)} & 2.5 \scalebox{.7}[1.0]{(2.4--2.5)} & 2.3 \scalebox{.7}[1.0]{(2.2--2.3)} \\
\textsc{iii} & 4.0 \scalebox{.7}[1.0]{(4.0--4.1)} & 4.0 \scalebox{.7}[1.0]{(3.9--4.0)} & 4.0 \scalebox{.7}[1.0]{(3.9--4.0)} & 3.2 \scalebox{.7}[1.0]{(3.2--3.3)} & 2.2 \scalebox{.7}[1.0]{(2.2--2.2)} & 1.8 \scalebox{.7}[1.0]{(1.8--1.9)} & 4.8 \scalebox{.7}[1.0]{(4.8--4.9)} & 4.2 \scalebox{.7}[1.0]{(4.2--4.3)} & 4.7 \scalebox{.7}[1.0]{(4.6--4.7)} & 4.5 \scalebox{.7}[1.0]{(4.4--4.5)} & 2.6 \scalebox{.7}[1.0]{(2.5--2.6)} & 2.4 \scalebox{.7}[1.0]{(2.4--2.5)} \\
\textsc{iv} & 2.6 \scalebox{.7}[1.0]{(2.6--2.6)} & 2.6 \scalebox{.7}[1.0]{(2.6--2.6)} & 2.6 \scalebox{.7}[1.0]{(2.6--2.6)} & 2.6 \scalebox{.7}[1.0]{(2.6--2.6)} & 2.6 \scalebox{.7}[1.0]{(2.6--2.6)} & 2.6 \scalebox{.7}[1.0]{(2.6--2.6)} & 3.2 \scalebox{.7}[1.0]{(3.2--3.2)} & 3.2 \scalebox{.7}[1.0]{(3.1--3.2)} & 4.3 \scalebox{.7}[1.0]{(4.3--4.3)} & 4.3 \scalebox{.7}[1.0]{(4.3--4.3)} & 3.2 \scalebox{.7}[1.0]{(3.2--3.2)} & 4.3 \scalebox{.7}[1.0]{(4.3--4.3)} \\
\textsc{v} & 2.6 \scalebox{.7}[1.0]{(2.6--2.6)} & 2.6 \scalebox{.7}[1.0]{(2.6--2.6)} & 2.6 \scalebox{.7}[1.0]{(2.6--2.6)} & 2.3 \scalebox{.7}[1.0]{(2.3--2.4)} & 2.6 \scalebox{.7}[1.0]{(2.6--2.6)} & 2.3 \scalebox{.7}[1.0]{(2.3--2.4)} & 3.7 \scalebox{.7}[1.0]{(3.7--3.8)} & 3.2 \scalebox{.7}[1.0]{(3.2--3.2)} & 4.3 \scalebox{.7}[1.0]{(4.3--4.3)} & 4.3 \scalebox{.7}[1.0]{(4.2--4.3)} & 3.7 \scalebox{.7}[1.0]{(3.7--3.8)} & 4.3 \scalebox{.7}[1.0]{(4.2--4.3)} \\
\textsc{vi} & 8.9 \scalebox{.7}[1.0]{(7.7--10.1)} & 4.3 \scalebox{.7}[1.0]{(3.4--5.2)} & 4.3 \scalebox{.7}[1.0]{(3.4--5.2)} & 5.1 \scalebox{.7}[1.0]{(4.1--6.0)} & 4.3 \scalebox{.7}[1.0]{(3.4--5.2)} & 2.6 \scalebox{.7}[1.0]{(1.9--3.3)} & 8.1 \scalebox{.7}[1.0]{(6.9--9.3)} & 4.9 \scalebox{.7}[1.0]{(4.0--5.8)} & 4.9 \scalebox{.7}[1.0]{(4.0--5.8)} & 5.8 \scalebox{.7}[1.0]{(4.7--6.8)} & 4.8 \scalebox{.7}[1.0]{(3.8--5.7)} & 3.0 \scalebox{.7}[1.0]{(2.3--3.7)} \\
\textsc{vii} & 4.3 \scalebox{.7}[1.0]{(4.0--4.7)} & 3.9 \scalebox{.7}[1.0]{(3.6--4.2)} & 3.9 \scalebox{.7}[1.0]{(3.6--4.2)} & 2.7 \scalebox{.7}[1.0]{(2.5--3.0)} & 2.4 \scalebox{.7}[1.0]{(2.1--2.7)} & 1.4 \scalebox{.7}[1.0]{(1.3--1.6)} & 5.3 \scalebox{.7}[1.0]{(4.8--5.8)} & 4.2 \scalebox{.7}[1.0]{(3.9--4.5)} & 4.6 \scalebox{.7}[1.0]{(4.3--4.9)} & 3.6 \scalebox{.7}[1.0]{(3.3--3.9)} & 2.9 \scalebox{.7}[1.0]{(2.5--3.2)} & 1.9 \scalebox{.7}[1.0]{(1.6--2.1)} \\
\textsc{viii} & 4.5 \scalebox{.7}[1.0]{(4.1--4.9)} & 4.2 \scalebox{.7}[1.0]{(3.9--4.5)} & 4.2 \scalebox{.7}[1.0]{(3.9--4.5)} & 2.9 \scalebox{.7}[1.0]{(2.6--3.1)} & 2.7 \scalebox{.7}[1.0]{(2.3--3.0)} & 1.6 \scalebox{.7}[1.0]{(1.4--1.9)} & 5.5 \scalebox{.7}[1.0]{(5.1--5.9)} & 4.4 \scalebox{.7}[1.0]{(4.0--4.7)} & 4.8 \scalebox{.7}[1.0]{(4.4--5.1)} & 3.9 \scalebox{.7}[1.0]{(3.6--4.2)} & 3.1 \scalebox{.7}[1.0]{(2.7--3.4)} & 2.1 \scalebox{.7}[1.0]{(1.8--2.3)} \\
\textsc{ix} & 3.8 \scalebox{.7}[1.0]{(3.6--4.0)} & 2.7 \scalebox{.7}[1.0]{(2.6--2.9)} & 2.7 \scalebox{.7}[1.0]{(2.6--2.9)} & 2.9 \scalebox{.7}[1.0]{(2.8--3.1)} & 3.7 \scalebox{.7}[1.0]{(3.5--4.0)} & 2.9 \scalebox{.7}[1.0]{(2.8--3.1)} & 4.6 \scalebox{.7}[1.0]{(4.3--4.9)} & 3.1 \scalebox{.7}[1.0]{(3.0--3.3)} & 4.2 \scalebox{.7}[1.0]{(4.0--4.4)} & 4.4 \scalebox{.7}[1.0]{(4.2--4.6)} & 4.6 \scalebox{.7}[1.0]{(4.3--4.8)} & 4.4 \scalebox{.7}[1.0]{(4.2--4.6)} \\
\textsc{x} & 3.4 \scalebox{.7}[1.0]{(3.2--3.7)} & 2.7 \scalebox{.7}[1.0]{(2.6--2.8)} & 2.7 \scalebox{.7}[1.0]{(2.6--2.8)} & 2.6 \scalebox{.7}[1.0]{(2.4--2.8)} & 3.4 \scalebox{.7}[1.0]{(3.2--3.6)} & 2.6 \scalebox{.7}[1.0]{(2.4--2.8)} & 5.1 \scalebox{.7}[1.0]{(4.8--5.4)} & 3.2 \scalebox{.7}[1.0]{(3.1--3.4)} & 4.3 \scalebox{.7}[1.0]{(4.1--4.5)} & 4.4 \scalebox{.7}[1.0]{(4.2--4.6)} & 5.0 \scalebox{.7}[1.0]{(4.8--5.3)} & 4.4 \scalebox{.7}[1.0]{(4.2--4.6)} \\
\midrule
\multicolumn{9}{l}{\textit{Positive side:}}\\
\textsc{i} & 4.8 \scalebox{.7}[1.0]{(3.8--5.7)} & 2.8 \scalebox{.7}[1.0]{(2.1--3.5)} & 2.8 \scalebox{.7}[1.0]{(2.1--3.5)} & 4.7 \scalebox{.7}[1.0]{(3.8--5.6)} & 2.8 \scalebox{.7}[1.0]{(2.1--3.5)} & 2.8 \scalebox{.7}[1.0]{(2.1--3.5)} & 5.1 \scalebox{.7}[1.0]{(4.2--6.1)} & 2.9 \scalebox{.7}[1.0]{(2.2--3.7)} & 2.9 \scalebox{.7}[1.0]{(2.2--3.7)} & 5.1 \scalebox{.7}[1.0]{(4.2--6.1)} & 2.9 \scalebox{.7}[1.0]{(2.2--3.7)} & 2.9 \scalebox{.7}[1.0]{(2.2--3.7)} \\
\textsc{ii} & 3.8 \scalebox{.7}[1.0]{(3.7--3.8)} & 1.9 \scalebox{.7}[1.0]{(1.9--2.0)} & 1.9 \scalebox{.7}[1.0]{(1.9--2.0)} & 3.0 \scalebox{.7}[1.0]{(3.0--3.0)} & 2.0 \scalebox{.7}[1.0]{(1.9--2.0)} & 1.6 \scalebox{.7}[1.0]{(1.6--1.6)} & 4.6 \scalebox{.7}[1.0]{(4.6--4.7)} & 2.1 \scalebox{.7}[1.0]{(2.0--2.1)} & 2.3 \scalebox{.7}[1.0]{(2.3--2.3)} & 4.3 \scalebox{.7}[1.0]{(4.3--4.3)} & 2.4 \scalebox{.7}[1.0]{(2.4--2.5)} & 2.3 \scalebox{.7}[1.0]{(2.2--2.3)} \\
\textsc{iii} & 4.7 \scalebox{.7}[1.0]{(3.8--5.6)} & 99.9 \scalebox{.7}[1.0]{(99.7--100.0)} & 99.9 \scalebox{.7}[1.0]{(99.7--100.0)} & 5.5 \scalebox{.7}[1.0]{(4.5--6.5)} & 4.7 \scalebox{.7}[1.0]{(3.8--5.6)} & 5.5 \scalebox{.7}[1.0]{(4.5--6.5)} & 3.2 \scalebox{.7}[1.0]{(2.5--4.0)} & 99.9 \scalebox{.7}[1.0]{(99.7--100.0)} & 100.0 \scalebox{.7}[1.0]{(99.9--100.0)} & 4.3 \scalebox{.7}[1.0]{(3.4--5.2)} & 3.2 \scalebox{.7}[1.0]{(2.5--4.0)} & 4.3 \scalebox{.7}[1.0]{(3.4--5.2)} \\
\textsc{iv} & 2.5 \scalebox{.7}[1.0]{(2.5--2.5)} & 2.5 \scalebox{.7}[1.0]{(2.5--2.5)} & 2.5 \scalebox{.7}[1.0]{(2.5--2.5)} & 2.5 \scalebox{.7}[1.0]{(2.5--2.5)} & 2.5 \scalebox{.7}[1.0]{(2.5--2.5)} & 2.5 \scalebox{.7}[1.0]{(2.5--2.5)} & 3.1 \scalebox{.7}[1.0]{(3.0--3.1)} & 3.1 \scalebox{.7}[1.0]{(3.0--3.1)} & 4.2 \scalebox{.7}[1.0]{(4.1--4.2)} & 4.2 \scalebox{.7}[1.0]{(4.1--4.2)} & 3.1 \scalebox{.7}[1.0]{(3.0--3.1)} & 4.2 \scalebox{.7}[1.0]{(4.1--4.2)} \\
\textsc{v} & 2.6 \scalebox{.7}[1.0]{(2.5--2.6)} & 6.0 \scalebox{.7}[1.0]{(5.9--6.1)} & 6.0 \scalebox{.7}[1.0]{(5.9--6.1)} & 3.6 \scalebox{.7}[1.0]{(3.5--3.7)} & 2.6 \scalebox{.7}[1.0]{(2.5--2.6)} & 3.6 \scalebox{.7}[1.0]{(3.5--3.7)} & 2.7 \scalebox{.7}[1.0]{(2.6--2.7)} & 7.4 \scalebox{.7}[1.0]{(7.3--7.4)} & 9.9 \scalebox{.7}[1.0]{(9.8--10.0)} & 4.5 \scalebox{.7}[1.0]{(4.4--4.5)} & 2.7 \scalebox{.7}[1.0]{(2.6--2.7)} & 4.5 \scalebox{.7}[1.0]{(4.4--4.5)} \\
\textsc{vi} & 4.4 \scalebox{.7}[1.0]{(3.5--5.3)} & 2.1 \scalebox{.7}[1.0]{(1.5--2.8)} & 2.1 \scalebox{.7}[1.0]{(1.5--2.8)} & 2.5 \scalebox{.7}[1.0]{(1.9--3.2)} & 2.1 \scalebox{.7}[1.0]{(1.5--2.8)} & 1.6 \scalebox{.7}[1.0]{(1.0--2.2)} & 3.8 \scalebox{.7}[1.0]{(2.9--4.6)} & 2.2 \scalebox{.7}[1.0]{(1.6--2.9)} & 2.2 \scalebox{.7}[1.0]{(1.6--2.9)} & 2.5 \scalebox{.7}[1.0]{(1.9--3.2)} & 2.1 \scalebox{.7}[1.0]{(1.5--2.8)} & 1.2 \scalebox{.7}[1.0]{(0.7--1.7)} \\
\textsc{vii} & 3.9 \scalebox{.7}[1.0]{(3.6--4.2)} & 2.0 \scalebox{.7}[1.0]{(1.8--2.1)} & 2.0 \scalebox{.7}[1.0]{(1.8--2.1)} & 2.5 \scalebox{.7}[1.0]{(2.3--2.7)} & 2.0 \scalebox{.7}[1.0]{(1.8--2.1)} & 1.2 \scalebox{.7}[1.0]{(1.1--1.3)} & 4.8 \scalebox{.7}[1.0]{(4.4--5.1)} & 2.0 \scalebox{.7}[1.0]{(1.8--2.2)} & 2.3 \scalebox{.7}[1.0]{(2.1--2.6)} & 3.3 \scalebox{.7}[1.0]{(3.1--3.6)} & 2.4 \scalebox{.7}[1.0]{(2.1--2.6)} & 1.6 \scalebox{.7}[1.0]{(1.4--1.7)} \\
\textsc{viii} & 2.7 \scalebox{.7}[1.0]{(2.0--3.4)} & 69.7 \scalebox{.7}[1.0]{(67.6--71.7)} & 69.7 \scalebox{.7}[1.0]{(67.6--71.7)} & 2.1 \scalebox{.7}[1.0]{(1.5--2.8)} & 2.7 \scalebox{.7}[1.0]{(2.0--3.4)} & 2.1 \scalebox{.7}[1.0]{(1.5--2.8)} & 2.7 \scalebox{.7}[1.0]{(2.0--3.4)} & 74.1 \scalebox{.7}[1.0]{(72.1--76.0)} & 77.5 \scalebox{.7}[1.0]{(75.7--79.3)} & 2.5 \scalebox{.7}[1.0]{(1.8--3.2)} & 2.7 \scalebox{.7}[1.0]{(2.0--3.4)} & 2.5 \scalebox{.7}[1.0]{(1.8--3.2)} \\
\textsc{ix} & 2.4 \scalebox{.7}[1.0]{(2.2--2.6)} & 1.9 \scalebox{.7}[1.0]{(1.8--2.1)} & 1.9 \scalebox{.7}[1.0]{(1.8--2.1)} & 2.0 \scalebox{.7}[1.0]{(1.9--2.2)} & 2.4 \scalebox{.7}[1.0]{(2.2--2.6)} & 2.0 \scalebox{.7}[1.0]{(1.9--2.1)} & 3.1 \scalebox{.7}[1.0]{(2.9--3.3)} & 2.4 \scalebox{.7}[1.0]{(2.2--2.5)} & 3.3 \scalebox{.7}[1.0]{(3.1--3.5)} & 3.4 \scalebox{.7}[1.0]{(3.2--3.6)} & 3.1 \scalebox{.7}[1.0]{(2.9--3.3)} & 3.4 \scalebox{.7}[1.0]{(3.2--3.6)} \\
\textsc{x} & 2.4 \scalebox{.7}[1.0]{(2.2--2.6)} & 3.8 \scalebox{.7}[1.0]{(3.6--4.0)} & 3.8 \scalebox{.7}[1.0]{(3.6--4.0)} & 2.6 \scalebox{.7}[1.0]{(2.4--2.9)} & 2.4 \scalebox{.7}[1.0]{(2.2--2.6)} & 2.6 \scalebox{.7}[1.0]{(2.4--2.8)} & 2.6 \scalebox{.7}[1.0]{(2.4--2.9)} & 4.6 \scalebox{.7}[1.0]{(4.3--4.8)} & 6.2 \scalebox{.7}[1.0]{(5.9--6.4)} & 3.3 \scalebox{.7}[1.0]{(3.0--3.5)} & 2.6 \scalebox{.7}[1.0]{(2.4--2.9)} & 3.3 \scalebox{.7}[1.0]{(3.0--3.5)} \\
\midrule
\multicolumn{9}{l}{\textit{Negative side:}}\\
\textsc{i} & 4.6 \scalebox{.7}[1.0]{(3.7--5.5)} & 2.7 \scalebox{.7}[1.0]{(2.0--3.4)} & 2.7 \scalebox{.7}[1.0]{(2.0--3.4)} & 4.7 \scalebox{.7}[1.0]{(3.7--5.6)} & 2.7 \scalebox{.7}[1.0]{(2.0--3.4)} & 2.7 \scalebox{.7}[1.0]{(2.0--3.4)} & 4.5 \scalebox{.7}[1.0]{(3.5--5.4)} & 2.2 \scalebox{.7}[1.0]{(1.6--2.9)} & 2.2 \scalebox{.7}[1.0]{(1.6--2.9)} & 4.4 \scalebox{.7}[1.0]{(3.5--5.3)} & 2.1 \scalebox{.7}[1.0]{(1.5--2.7)} & 2.2 \scalebox{.7}[1.0]{(1.6--2.8)} \\
\textsc{ii} & 4.2 \scalebox{.7}[1.0]{(3.3--5.1)} & 99.6 \scalebox{.7}[1.0]{(99.3--99.9)} & 99.6 \scalebox{.7}[1.0]{(99.3--99.9)} & 5.1 \scalebox{.7}[1.0]{(4.1--6.1)} & 4.2 \scalebox{.7}[1.0]{(3.3--5.1)} & 5.1 \scalebox{.7}[1.0]{(4.1--6.1)} & 5.0 \scalebox{.7}[1.0]{(4.0--5.9)} & 100.0 \scalebox{.7}[1.0]{(99.9--100.0)} & 100.0 \scalebox{.7}[1.0]{(100.0--100.0)} & 6.5 \scalebox{.7}[1.0]{(5.4--7.6)} & 5.0 \scalebox{.7}[1.0]{(4.0--5.9)} & 6.5 \scalebox{.7}[1.0]{(5.4--7.6)} \\
\textsc{iii} & 4.0 \scalebox{.7}[1.0]{(4.0--4.0)} & 2.1 \scalebox{.7}[1.0]{(2.1--2.2)} & 2.1 \scalebox{.7}[1.0]{(2.1--2.2)} & 3.2 \scalebox{.7}[1.0]{(3.2--3.3)} & 2.2 \scalebox{.7}[1.0]{(2.1--2.2)} & 1.8 \scalebox{.7}[1.0]{(1.8--1.8)} & 4.8 \scalebox{.7}[1.0]{(4.8--4.8)} & 2.2 \scalebox{.7}[1.0]{(2.2--2.3)} & 2.5 \scalebox{.7}[1.0]{(2.4--2.5)} & 4.5 \scalebox{.7}[1.0]{(4.4--4.5)} & 2.6 \scalebox{.7}[1.0]{(2.5--2.6)} & 2.4 \scalebox{.7}[1.0]{(2.4--2.4)} \\
\textsc{iv} & 2.7 \scalebox{.7}[1.0]{(2.7--2.7)} & 2.7 \scalebox{.7}[1.0]{(2.7--2.7)} & 2.7 \scalebox{.7}[1.0]{(2.7--2.7)} & 2.7 \scalebox{.7}[1.0]{(2.7--2.7)} & 2.7 \scalebox{.7}[1.0]{(2.7--2.7)} & 2.7 \scalebox{.7}[1.0]{(2.7--2.7)} & 3.3 \scalebox{.7}[1.0]{(3.2--3.3)} & 3.3 \scalebox{.7}[1.0]{(3.2--3.3)} & 4.4 \scalebox{.7}[1.0]{(4.3--4.4)} & 4.4 \scalebox{.7}[1.0]{(4.3--4.4)} & 3.3 \scalebox{.7}[1.0]{(3.2--3.3)} & 4.4 \scalebox{.7}[1.0]{(4.3--4.4)} \\
\textsc{v} & 2.6 \scalebox{.7}[1.0]{(2.6--2.6)} & 1.7 \scalebox{.7}[1.0]{(1.7--1.7)} & 1.7 \scalebox{.7}[1.0]{(1.7--1.7)} & 2.0 \scalebox{.7}[1.0]{(2.0--2.1)} & 2.6 \scalebox{.7}[1.0]{(2.6--2.6)} & 2.0 \scalebox{.7}[1.0]{(2.0--2.1)} & 3.9 \scalebox{.7}[1.0]{(3.9--3.9)} & 2.1 \scalebox{.7}[1.0]{(2.1--2.1)} & 2.8 \scalebox{.7}[1.0]{(2.8--2.8)} & 4.2 \scalebox{.7}[1.0]{(4.2--4.2)} & 3.9 \scalebox{.7}[1.0]{(3.9--3.9)} & 4.2 \scalebox{.7}[1.0]{(4.2--4.2)} \\
\textsc{vi} & 4.5 \scalebox{.7}[1.0]{(3.6--5.4)} & 2.1 \scalebox{.7}[1.0]{(1.5--2.8)} & 2.1 \scalebox{.7}[1.0]{(1.5--2.8)} & 2.5 \scalebox{.7}[1.0]{(1.8--3.2)} & 2.1 \scalebox{.7}[1.0]{(1.5--2.8)} & 1.0 \scalebox{.7}[1.0]{(0.6--1.4)} & 4.3 \scalebox{.7}[1.0]{(3.5--5.2)} & 2.6 \scalebox{.7}[1.0]{(1.9--3.4)} & 2.6 \scalebox{.7}[1.0]{(1.9--3.4)} & 3.2 \scalebox{.7}[1.0]{(2.4--4.0)} & 2.6 \scalebox{.7}[1.0]{(1.9--3.3)} & 1.8 \scalebox{.7}[1.0]{(1.2--2.4)} \\
\textsc{vii} & 2.6 \scalebox{.7}[1.0]{(1.9--3.3)} & 68.6 \scalebox{.7}[1.0]{(66.6--70.6)} & 68.6 \scalebox{.7}[1.0]{(66.6--70.6)} & 2.2 \scalebox{.7}[1.0]{(1.6--2.9)} & 2.6 \scalebox{.7}[1.0]{(1.9--3.3)} & 2.2 \scalebox{.7}[1.0]{(1.6--2.9)} & 3.2 \scalebox{.7}[1.0]{(2.5--4.0)} & 73.9 \scalebox{.7}[1.0]{(72.0--75.8)} & 77.2 \scalebox{.7}[1.0]{(75.4--79.1)} & 2.9 \scalebox{.7}[1.0]{(2.2--3.6)} & 3.2 \scalebox{.7}[1.0]{(2.5--4.0)} & 2.9 \scalebox{.7}[1.0]{(2.2--3.6)} \\
\textsc{viii} & 3.9 \scalebox{.7}[1.0]{(3.6--4.2)} & 2.1 \scalebox{.7}[1.0]{(1.9--2.3)} & 2.1 \scalebox{.7}[1.0]{(1.9--2.3)} & 2.6 \scalebox{.7}[1.0]{(2.4--2.8)} & 2.1 \scalebox{.7}[1.0]{(1.9--2.3)} & 1.4 \scalebox{.7}[1.0]{(1.3--1.5)} & 5.0 \scalebox{.7}[1.0]{(4.6--5.3)} & 2.2 \scalebox{.7}[1.0]{(2.0--2.4)} & 2.5 \scalebox{.7}[1.0]{(2.3--2.7)} & 3.5 \scalebox{.7}[1.0]{(3.3--3.7)} & 2.5 \scalebox{.7}[1.0]{(2.3--2.7)} & 1.7 \scalebox{.7}[1.0]{(1.6--1.9)} \\
\textsc{ix} & 2.8 \scalebox{.7}[1.0]{(2.6--3.0)} & 2.2 \scalebox{.7}[1.0]{(2.1--2.4)} & 2.2 \scalebox{.7}[1.0]{(2.1--2.4)} & 2.3 \scalebox{.7}[1.0]{(2.2--2.5)} & 2.8 \scalebox{.7}[1.0]{(2.6--2.9)} & 2.3 \scalebox{.7}[1.0]{(2.2--2.5)} & 3.3 \scalebox{.7}[1.0]{(3.1--3.6)} & 2.6 \scalebox{.7}[1.0]{(2.4--2.7)} & 3.5 \scalebox{.7}[1.0]{(3.4--3.7)} & 3.6 \scalebox{.7}[1.0]{(3.4--3.8)} & 3.3 \scalebox{.7}[1.0]{(3.1--3.6)} & 3.6 \scalebox{.7}[1.0]{(3.4--3.8)} \\
\textsc{x} & 2.6 \scalebox{.7}[1.0]{(2.4--2.8)} & 1.6 \scalebox{.7}[1.0]{(1.5--1.7)} & 1.6 \scalebox{.7}[1.0]{(1.5--1.7)} & 1.8 \scalebox{.7}[1.0]{(1.7--1.9)} & 2.6 \scalebox{.7}[1.0]{(2.4--2.7)} & 1.8 \scalebox{.7}[1.0]{(1.7--1.9)} & 4.3 \scalebox{.7}[1.0]{(4.0--4.5)} & 2.0 \scalebox{.7}[1.0]{(1.8--2.1)} & 2.8 \scalebox{.7}[1.0]{(2.6--3.0)} & 3.7 \scalebox{.7}[1.0]{(3.5--3.9)} & 4.2 \scalebox{.7}[1.0]{(4.0--4.5)} & 3.7 \scalebox{.7}[1.0]{(3.5--3.9)} \\
\bottomrule
\end{tabular}}
\end{center}
\label{tab:synthetic_fdp_noref}
\end{sidewaystable}

\begin{sidewaystable}[!p]
\setcounter{sidewaystablepage}{3}
\begin{center}
{\scriptsize
\hspace*{-24mm}
\begin{tabular}{@{}
p{10mm}@{\hspace{1mm}}
p{20mm}@{\hspace{1mm}}
p{20mm}@{\hspace{1mm}}
p{20mm}@{\hspace{1mm}}
p{20mm}@{\hspace{1mm}}
p{20mm}@{\hspace{1mm}}
p{20mm}@{\hspace{1mm}}
p{20mm}@{\hspace{1mm}}
p{20mm}@{\hspace{1mm}}
p{20mm}@{\hspace{1mm}}
p{20mm}@{\hspace{1mm}}
p{20mm}@{\hspace{1mm}}
p{20mm}@{}}
\toprule
{} & 
\multicolumn{6}{c}{Benjamini--Hochberg} & 
\multicolumn{6}{c}{Benjamini--Krieger--Yekutieli} \\
\cmidrule(r){2-7}\cmidrule(){8-13}
Scn. & 
Canonical & Combined & Two-tailed & Split-tails & Canonical+\textsc{bb} & Split-tails+\textsc{bb} &
Canonical & Combined & Two-tailed & Split-tails & Canonical+\textsc{bb} & Split-tails+\textsc{bb} \\
\midrule
\multicolumn{9}{l}{\textit{Both sides:}}\\
\textsc{i} & --- & --- & --- & --- & --- & --- & --- & --- & --- & --- & --- & --- \\
\textsc{ii} & 77.5 \scalebox{.7}[1.0]{(77.4--77.6)} & 66.6 \scalebox{.7}[1.0]{(66.5--66.8)} & 66.6 \scalebox{.7}[1.0]{(66.5--66.8)} & 73.5 \scalebox{.7}[1.0]{(73.4--73.6)} & 65.6 \scalebox{.7}[1.0]{(65.4--65.8)} & 61.8 \scalebox{.7}[1.0]{(61.6--62.0)} & 81.5 \scalebox{.7}[1.0]{(81.4--81.6)} & 68.3 \scalebox{.7}[1.0]{(68.2--68.4)} & 70.3 \scalebox{.7}[1.0]{(70.2--70.4)} & 80.1 \scalebox{.7}[1.0]{(80.0--80.2)} & 69.4 \scalebox{.7}[1.0]{(69.2--69.6)} & 67.9 \scalebox{.7}[1.0]{(67.6--68.1)} \\
\textsc{iii} & 77.8 \scalebox{.7}[1.0]{(77.6--77.9)} & 66.8 \scalebox{.7}[1.0]{(66.7--66.9)} & 66.8 \scalebox{.7}[1.0]{(66.7--66.9)} & 73.6 \scalebox{.7}[1.0]{(73.5--73.8)} & 65.8 \scalebox{.7}[1.0]{(65.6--66.0)} & 62.0 \scalebox{.7}[1.0]{(61.8--62.1)} & 81.7 \scalebox{.7}[1.0]{(81.6--81.8)} & 68.5 \scalebox{.7}[1.0]{(68.4--68.6)} & 70.5 \scalebox{.7}[1.0]{(70.3--70.6)} & 80.3 \scalebox{.7}[1.0]{(80.2--80.4)} & 69.3 \scalebox{.7}[1.0]{(69.1--69.5)} & 67.9 \scalebox{.7}[1.0]{(67.7--68.1)} \\
\textsc{iv} & 76.4 \scalebox{.7}[1.0]{(76.4--76.5)} & 76.4 \scalebox{.7}[1.0]{(76.4--76.5)} & 76.4 \scalebox{.7}[1.0]{(76.4--76.5)} & 76.4 \scalebox{.7}[1.0]{(76.4--76.5)} & 76.4 \scalebox{.7}[1.0]{(76.4--76.5)} & 76.4 \scalebox{.7}[1.0]{(76.4--76.5)} & 80.0 \scalebox{.7}[1.0]{(79.9--80.1)} & 80.0 \scalebox{.7}[1.0]{(79.9--80.1)} & 85.3 \scalebox{.7}[1.0]{(85.2--85.4)} & 85.3 \scalebox{.7}[1.0]{(85.2--85.4)} & 80.0 \scalebox{.7}[1.0]{(79.9--80.1)} & 85.3 \scalebox{.7}[1.0]{(85.2--85.4)} \\
\textsc{v} & 78.8 \scalebox{.7}[1.0]{(78.7--78.8)} & 76.4 \scalebox{.7}[1.0]{(76.3--76.5)} & 76.4 \scalebox{.7}[1.0]{(76.3--76.5)} & 76.8 \scalebox{.7}[1.0]{(76.8--76.9)} & 78.8 \scalebox{.7}[1.0]{(78.7--78.8)} & 76.8 \scalebox{.7}[1.0]{(76.8--76.9)} & 84.1 \scalebox{.7}[1.0]{(84.0--84.2)} & 80.0 \scalebox{.7}[1.0]{(79.9--80.1)} & 85.3 \scalebox{.7}[1.0]{(85.2--85.4)} & 87.0 \scalebox{.7}[1.0]{(86.9--87.1)} & 84.1 \scalebox{.7}[1.0]{(84.0--84.2)} & 87.0 \scalebox{.7}[1.0]{(86.9--87.1)} \\
\textsc{vi} & --- & --- & --- & --- & --- & --- & --- & --- & --- & --- & --- & --- \\
\textsc{vii} & 77.6 \scalebox{.7}[1.0]{(76.5--78.8)} & 65.3 \scalebox{.7}[1.0]{(64.1--66.4)} & 65.3 \scalebox{.7}[1.0]{(64.1--66.4)} & 74.1 \scalebox{.7}[1.0]{(73.2--75.1)} & 64.8 \scalebox{.7}[1.0]{(63.6--66.0)} & 61.5 \scalebox{.7}[1.0]{(60.5--62.5)} & 80.7 \scalebox{.7}[1.0]{(79.2--82.2)} & 65.0 \scalebox{.7}[1.0]{(63.8--66.2)} & 66.9 \scalebox{.7}[1.0]{(65.6--68.2)} & 79.7 \scalebox{.7}[1.0]{(78.6--80.7)} & 66.7 \scalebox{.7}[1.0]{(65.4--68.0)} & 65.6 \scalebox{.7}[1.0]{(64.6--66.6)} \\
\textsc{viii} & 77.0 \scalebox{.7}[1.0]{(75.8--78.2)} & 64.8 \scalebox{.7}[1.0]{(63.7--66.0)} & 64.8 \scalebox{.7}[1.0]{(63.7--66.0)} & 73.7 \scalebox{.7}[1.0]{(72.7--74.7)} & 64.3 \scalebox{.7}[1.0]{(63.1--65.5)} & 61.1 \scalebox{.7}[1.0]{(60.1--62.1)} & 81.1 \scalebox{.7}[1.0]{(79.7--82.4)} & 65.6 \scalebox{.7}[1.0]{(64.5--66.8)} & 67.5 \scalebox{.7}[1.0]{(66.3--68.7)} & 80.2 \scalebox{.7}[1.0]{(79.2--81.1)} & 67.2 \scalebox{.7}[1.0]{(65.9--68.4)} & 66.1 \scalebox{.7}[1.0]{(65.1--67.1)} \\
\textsc{ix} & 74.6 \scalebox{.7}[1.0]{(74.4--74.8)} & 76.3 \scalebox{.7}[1.0]{(76.1--76.4)} & 76.3 \scalebox{.7}[1.0]{(76.1--76.4)} & 75.8 \scalebox{.7}[1.0]{(75.6--75.9)} & 74.6 \scalebox{.7}[1.0]{(74.4--74.8)} & 75.7 \scalebox{.7}[1.0]{(75.6--75.9)} & 78.5 \scalebox{.7}[1.0]{(78.3--78.6)} & 80.0 \scalebox{.7}[1.0]{(79.9--80.1)} & 85.2 \scalebox{.7}[1.0]{(85.1--85.4)} & 85.1 \scalebox{.7}[1.0]{(84.9--85.2)} & 78.4 \scalebox{.7}[1.0]{(78.2--78.6)} & 85.1 \scalebox{.7}[1.0]{(84.9--85.2)} \\
\textsc{x} & 77.2 \scalebox{.7}[1.0]{(76.7--77.8)} & 75.6 \scalebox{.7}[1.0]{(75.0--76.2)} & 75.6 \scalebox{.7}[1.0]{(75.0--76.2)} & 76.1 \scalebox{.7}[1.0]{(75.6--76.6)} & 77.2 \scalebox{.7}[1.0]{(76.6--77.7)} & 76.1 \scalebox{.7}[1.0]{(75.6--76.6)} & 82.7 \scalebox{.7}[1.0]{(82.1--83.2)} & 79.1 \scalebox{.7}[1.0]{(78.6--79.7)} & 84.1 \scalebox{.7}[1.0]{(83.6--84.7)} & 86.4 \scalebox{.7}[1.0]{(86.0--86.9)} & 82.6 \scalebox{.7}[1.0]{(82.0--83.1)} & 86.4 \scalebox{.7}[1.0]{(86.0--86.9)} \\
\midrule
\multicolumn{9}{l}{\textit{Positive side:}}\\
\textsc{i} & --- & --- & --- & --- & --- & --- & --- & --- & --- & --- & --- & --- \\
\textsc{ii} & 77.6 \scalebox{.7}[1.0]{(77.5--77.7)} & 65.5 \scalebox{.7}[1.0]{(65.4--65.6)} & 65.5 \scalebox{.7}[1.0]{(65.4--65.6)} & 73.6 \scalebox{.7}[1.0]{(73.5--73.7)} & 65.7 \scalebox{.7}[1.0]{(65.5--65.8)} & 61.9 \scalebox{.7}[1.0]{(61.7--62.0)} & 81.6 \scalebox{.7}[1.0]{(81.5--81.7)} & 67.0 \scalebox{.7}[1.0]{(66.9--67.1)} & 68.8 \scalebox{.7}[1.0]{(68.7--68.9)} & 80.2 \scalebox{.7}[1.0]{(80.1--80.3)} & 69.5 \scalebox{.7}[1.0]{(69.3--69.7)} & 67.9 \scalebox{.7}[1.0]{(67.7--68.2)} \\
\textsc{iii} & --- & --- & --- & --- & --- & --- & --- & --- & --- & --- & --- & --- \\
\textsc{iv} & 76.5 \scalebox{.7}[1.0]{(76.4--76.6)} & 76.5 \scalebox{.7}[1.0]{(76.4--76.6)} & 76.5 \scalebox{.7}[1.0]{(76.4--76.6)} & 76.5 \scalebox{.7}[1.0]{(76.4--76.6)} & 76.5 \scalebox{.7}[1.0]{(76.4--76.6)} & 76.5 \scalebox{.7}[1.0]{(76.4--76.6)} & 80.1 \scalebox{.7}[1.0]{(79.9--80.2)} & 80.1 \scalebox{.7}[1.0]{(80.0--80.2)} & 85.3 \scalebox{.7}[1.0]{(85.2--85.4)} & 85.3 \scalebox{.7}[1.0]{(85.2--85.4)} & 80.1 \scalebox{.7}[1.0]{(79.9--80.2)} & 85.3 \scalebox{.7}[1.0]{(85.2--85.4)} \\
\textsc{v} & 62.2 \scalebox{.7}[1.0]{(61.9--62.4)} & 79.4 \scalebox{.7}[1.0]{(79.2--79.5)} & 79.4 \scalebox{.7}[1.0]{(79.2--79.5)} & 68.9 \scalebox{.7}[1.0]{(68.7--69.1)} & 62.2 \scalebox{.7}[1.0]{(61.9--62.4)} & 68.9 \scalebox{.7}[1.0]{(68.7--69.1)} & 63.0 \scalebox{.7}[1.0]{(62.8--63.2)} & 83.6 \scalebox{.7}[1.0]{(83.5--83.8)} & 90.6 \scalebox{.7}[1.0]{(90.5--90.8)} & 73.0 \scalebox{.7}[1.0]{(72.8--73.2)} & 63.0 \scalebox{.7}[1.0]{(62.8--63.2)} & 73.0 \scalebox{.7}[1.0]{(72.8--73.2)} \\
\textsc{vi} & --- & --- & --- & --- & --- & --- & --- & --- & --- & --- & --- & --- \\
\textsc{vii} & 77.6 \scalebox{.7}[1.0]{(76.4--78.7)} & 64.8 \scalebox{.7}[1.0]{(63.6--66.0)} & 64.8 \scalebox{.7}[1.0]{(63.6--66.0)} & 74.2 \scalebox{.7}[1.0]{(73.2--75.1)} & 64.8 \scalebox{.7}[1.0]{(63.6--65.9)} & 61.5 \scalebox{.7}[1.0]{(60.5--62.5)} & 80.6 \scalebox{.7}[1.0]{(79.1--82.1)} & 64.4 \scalebox{.7}[1.0]{(63.2--65.7)} & 66.3 \scalebox{.7}[1.0]{(65.0--67.6)} & 79.7 \scalebox{.7}[1.0]{(78.6--80.7)} & 66.6 \scalebox{.7}[1.0]{(65.3--67.9)} & 65.6 \scalebox{.7}[1.0]{(64.5--66.6)} \\
\textsc{viii} & --- & --- & --- & --- & --- & --- & --- & --- & --- & --- & --- & --- \\
\textsc{ix} & 74.2 \scalebox{.7}[1.0]{(73.0--75.3)} & 76.0 \scalebox{.7}[1.0]{(75.2--76.9)} & 76.0 \scalebox{.7}[1.0]{(75.2--76.9)} & 75.4 \scalebox{.7}[1.0]{(74.5--76.4)} & 74.1 \scalebox{.7}[1.0]{(73.0--75.2)} & 75.4 \scalebox{.7}[1.0]{(74.5--76.4)} & 78.4 \scalebox{.7}[1.0]{(77.2--79.5)} & 80.0 \scalebox{.7}[1.0]{(79.2--80.7)} & 85.2 \scalebox{.7}[1.0]{(84.4--86.0)} & 84.9 \scalebox{.7}[1.0]{(84.0--85.8)} & 78.4 \scalebox{.7}[1.0]{(77.2--79.5)} & 84.9 \scalebox{.7}[1.0]{(84.0--85.8)} \\
\textsc{x} & 60.7 \scalebox{.7}[1.0]{(59.5--62.0)} & 78.6 \scalebox{.7}[1.0]{(77.8--79.5)} & 78.6 \scalebox{.7}[1.0]{(77.8--79.5)} & 68.5 \scalebox{.7}[1.0]{(67.5--69.5)} & 60.7 \scalebox{.7}[1.0]{(59.4--62.0)} & 68.5 \scalebox{.7}[1.0]{(67.4--69.5)} & 61.7 \scalebox{.7}[1.0]{(60.3--63.0)} & 82.7 \scalebox{.7}[1.0]{(81.8--83.5)} & 89.0 \scalebox{.7}[1.0]{(88.2--89.8)} & 73.0 \scalebox{.7}[1.0]{(71.9--74.0)} & 61.7 \scalebox{.7}[1.0]{(60.3--63.0)} & 73.0 \scalebox{.7}[1.0]{(71.9--74.0)} \\
\midrule
\multicolumn{9}{l}{\textit{Negative side:}}\\
\textsc{i} & --- & --- & --- & --- & --- & --- & --- & --- & --- & --- & --- & --- \\
\textsc{ii} & --- & --- & --- & --- & --- & --- & --- & --- & --- & --- & --- & --- \\
\textsc{iii} & 77.9 \scalebox{.7}[1.0]{(77.7--78.0)} & 65.6 \scalebox{.7}[1.0]{(65.5--65.8)} & 65.6 \scalebox{.7}[1.0]{(65.5--65.8)} & 73.7 \scalebox{.7}[1.0]{(73.6--73.9)} & 65.9 \scalebox{.7}[1.0]{(65.7--66.1)} & 62.0 \scalebox{.7}[1.0]{(61.9--62.2)} & 81.8 \scalebox{.7}[1.0]{(81.7--81.9)} & 67.2 \scalebox{.7}[1.0]{(67.1--67.3)} & 69.0 \scalebox{.7}[1.0]{(68.8--69.1)} & 80.4 \scalebox{.7}[1.0]{(80.3--80.5)} & 69.4 \scalebox{.7}[1.0]{(69.2--69.6)} & 68.0 \scalebox{.7}[1.0]{(67.8--68.2)} \\
\textsc{iv} & 76.6 \scalebox{.7}[1.0]{(76.5--76.7)} & 76.6 \scalebox{.7}[1.0]{(76.5--76.7)} & 76.6 \scalebox{.7}[1.0]{(76.5--76.7)} & 76.6 \scalebox{.7}[1.0]{(76.5--76.7)} & 76.6 \scalebox{.7}[1.0]{(76.5--76.7)} & 76.6 \scalebox{.7}[1.0]{(76.5--76.7)} & 80.2 \scalebox{.7}[1.0]{(80.1--80.3)} & 80.2 \scalebox{.7}[1.0]{(80.1--80.3)} & 85.5 \scalebox{.7}[1.0]{(85.4--85.6)} & 85.5 \scalebox{.7}[1.0]{(85.3--85.6)} & 80.2 \scalebox{.7}[1.0]{(80.1--80.3)} & 85.5 \scalebox{.7}[1.0]{(85.3--85.6)} \\
\textsc{v} & 83.0 \scalebox{.7}[1.0]{(83.0--83.1)} & 75.8 \scalebox{.7}[1.0]{(75.7--75.9)} & 75.8 \scalebox{.7}[1.0]{(75.7--75.9)} & 79.0 \scalebox{.7}[1.0]{(78.9--79.0)} & 83.0 \scalebox{.7}[1.0]{(83.0--83.1)} & 79.0 \scalebox{.7}[1.0]{(78.9--79.0)} & 89.5 \scalebox{.7}[1.0]{(89.5--89.6)} & 79.2 \scalebox{.7}[1.0]{(79.2--79.3)} & 84.1 \scalebox{.7}[1.0]{(84.0--84.2)} & 90.7 \scalebox{.7}[1.0]{(90.6--90.8)} & 89.5 \scalebox{.7}[1.0]{(89.5--89.6)} & 90.7 \scalebox{.7}[1.0]{(90.6--90.8)} \\
\textsc{vi} & --- & --- & --- & --- & --- & --- & --- & --- & --- & --- & --- & --- \\
\textsc{vii} & --- & --- & --- & --- & --- & --- & --- & --- & --- & --- & --- & --- \\
\textsc{viii} & 76.9 \scalebox{.7}[1.0]{(75.7--78.1)} & 64.3 \scalebox{.7}[1.0]{(63.2--65.5)} & 64.3 \scalebox{.7}[1.0]{(63.2--65.5)} & 73.7 \scalebox{.7}[1.0]{(72.7--74.7)} & 64.3 \scalebox{.7}[1.0]{(63.1--65.5)} & 61.1 \scalebox{.7}[1.0]{(60.1--62.1)} & 81.0 \scalebox{.7}[1.0]{(79.6--82.3)} & 65.1 \scalebox{.7}[1.0]{(63.9--66.3)} & 66.9 \scalebox{.7}[1.0]{(65.6--68.1)} & 80.2 \scalebox{.7}[1.0]{(79.2--81.1)} & 67.1 \scalebox{.7}[1.0]{(65.8--68.3)} & 66.1 \scalebox{.7}[1.0]{(65.1--67.1)} \\
\textsc{ix} & 75.1 \scalebox{.7}[1.0]{(74.0--76.2)} & 76.7 \scalebox{.7}[1.0]{(75.8--77.5)} & 76.7 \scalebox{.7}[1.0]{(75.8--77.5)} & 76.2 \scalebox{.7}[1.0]{(75.3--77.2)} & 75.1 \scalebox{.7}[1.0]{(74.0--76.2)} & 76.2 \scalebox{.7}[1.0]{(75.3--77.1)} & 78.6 \scalebox{.7}[1.0]{(77.5--79.7)} & 80.1 \scalebox{.7}[1.0]{(79.3--80.9)} & 85.4 \scalebox{.7}[1.0]{(84.6--86.2)} & 85.3 \scalebox{.7}[1.0]{(84.5--86.2)} & 78.6 \scalebox{.7}[1.0]{(77.5--79.7)} & 85.3 \scalebox{.7}[1.0]{(84.5--86.2)} \\
\textsc{x} & 81.4 \scalebox{.7}[1.0]{(80.5--82.4)} & 74.9 \scalebox{.7}[1.0]{(74.0--75.8)} & 74.9 \scalebox{.7}[1.0]{(74.0--75.8)} & 78.1 \scalebox{.7}[1.0]{(77.2--78.9)} & 81.3 \scalebox{.7}[1.0]{(80.4--82.3)} & 78.1 \scalebox{.7}[1.0]{(77.2--78.9)} & 88.0 \scalebox{.7}[1.0]{(87.0--89.0)} & 78.3 \scalebox{.7}[1.0]{(77.4--79.2)} & 83.0 \scalebox{.7}[1.0]{(82.1--83.9)} & 89.9 \scalebox{.7}[1.0]{(89.1--90.7)} & 87.9 \scalebox{.7}[1.0]{(86.9--88.9)} & 89.9 \scalebox{.7}[1.0]{(89.1--90.7)} \\
\bottomrule
\end{tabular}}
\end{center}
\label{tab:synthetic_pwr_noref}
\end{sidewaystable}

\subsubsection{Correction strategies}
\label{sec:results_strategies}

In the absence of signal, that is, under the complete null hypothesis, and with independent test statistics (scenario \textsc{i}), the canonical and split-tails methods worked well when false positives were counted among observed positive test statistics only, or among observed negative test statistics only. In these cases, the empirical false discovery proportion was within the confidence intervals, being neither conservative nor anti-conservative. When false positives were counted among all test statistics (i.e., both sides), the observed error rate was twice as large as as the nominal test level. Combined and two-tailed methods controlled the error rate at the nominal level, albeit became conservative when only one or the other direction was considered; these two methods, however, became invalid other scenarios. The \textsc{bb} adjustment with canonical and split-tails controlled the error rate in all scenarios with either \textsc{bh} or \textsc{bky}.

With only positive or only negative true signal (scenarios \textsc{ii} and \textsc{iii}), all six correction strategies produced error rates that were within the expected level for test statistics with the same side as the signal, even if leaning heavily towards conservativeness. However, for test statistics with the opposite sign as the signal, the observed proportion of false discoveries was not controlled by a large margin, approaching 100\% error rate, for the combined and two-tailed methods. When both sides were considered to count false positives (that is, all tests considered), the error rate was controlled within the nominal level, leaning again towards conservativeness, albeit less heavily.

With both positive and negative true signal in a balanced scenario (scenario \textsc{iv}), the error rate was controlled regardless of how the error rate was counted (positive test statistics only, negative only, or both). That control, however, leaned towards conservativeness. With unbalanced amounts of signal (scenario \textsc{v}), which more typically represents real experiments, the error rate was not controlled in the side opposite to the one with a preponderance of true signal, with the nominal error rate lying outside the confidence interval, while conservative on the same side.

\subsubsection{\textsc{fdr} methods}

While the same patterns above were observed for both \textsc{bh} and \textsc{bky} methods, \textsc{bky} had the advantage of leading to observed false discovery rates that were closer to the nominal level in the presence of signal with the same direction as the observed test statistic, as opposed to leaning towards conservatism as \textsc{bh}. In other words, whenever there was signal, \textsc{bky} was more powerful than \textsc{bh}. However, this also meant that the problems with excess false discoveries observed in the side opposite from where there was more signal was also exacerbated (scenario \textsc{v}).

\subsubsection{Selective inference}

Under the complete null (scenario \textsc{i}), canonical + \textsc{bb} and split-tails + \textsc{bb} controlled the global (both sides) false discovery rate at the nominal level, eliminating the doubled error rate, yet yielding identical results as their non-\textsc{bb} counterparts.

\subsubsection{Dependencies among tests}

The same patterns above were observed for dependent tests (scenarios \textsc{vi}--\textsc{x}). Introduced dependencies did not substantially alter the results when comparing the correction strategies or the different \textsc{fdr} methods.

\subsection{Real data}

Figure~\ref{fig:narps} shows maps thresholded at positive and negative sides using the different \textsc{fdr} methods and correction strategies, as well as with uncorrected p-values; the thresholds in each case ($t$ statistics) are also shown.

\begin{figure}[!b]
\caption{\textit{(page \pageref{fig:narps_noref})} Thresholded statistical maps using different \textsc{fdr} methods and correction strategies for directional inference, as well as uncorrected, for the \textsc{narps} data. Significant results in the positive side are those greater than the threshold $t_{\text{pos}}$; in the negative side, those smaller (more negative) than $t_{\text{neg}}$; if nothing is declared significant, the threshold is set, for display purposes, at a value more extreme than the test statistic at that side, here generically represented as $\infty$. The canonical and split-tails strategies produce generally similar results; likewise, combined and two-tailed methods produce broadly similar maps. Inferences using combined or two-tailed maps should not make directional statements about the results; these are only possible in the canonical and split-tails methods. The figure also shows slight but consistent gains in power with the \textsc{bky} procedure when compared to the usual \textsc{bh} procedure. Only visualization software that allow asymmetrical thresholds can be used to display \textsc{fdr}-corrected results that enable directional inferences.}
\label{fig:narps}
\end{figure}

\begin{figure}[!p]
\centering
\includegraphics[width=1\textwidth]{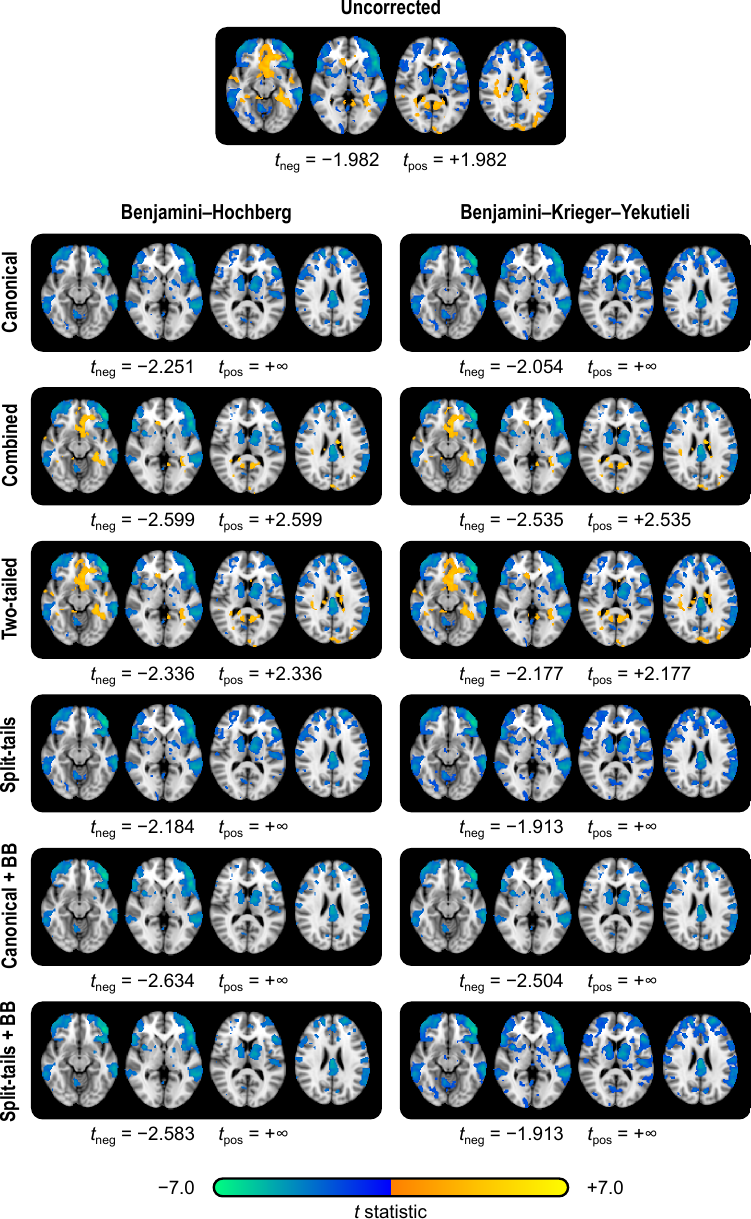}
\label{fig:narps_noref}
\end{figure}

The uncorrected map is thresholded, at the positive side, at $t_{\text{pos}} = 1 - t_{\text{cdf}}^{-1}(\alpha/2,\nu) = +1.982$, where $t_{\text{cdf}}^{-1}$ is the inverse cumulative distribution function (cdf) of the Student's $t$-distribution, $\nu=107$ are the degrees of freedom for this analysis, and $\alpha=0.05$ is the test level; it is thresholded similarly at the negative side at $t_{\text{neg}} = t_{\text{cdf}}^{-1}(\alpha/2,\nu) = -1.982$. The uncorrected thresholded maps suggests the presence of signal in both positive and negative directions, with a preponderance of effects in the negative side, involving more brain regions, particularly around the cortex.

Thresholds were calculated for the two two correction methods and six strategies, as shown in Figure~\ref{fig:narps}, using the inverse cdf of Student's $t$-distribution with the same $\nu=107$, but with test level determined by the threshold computed from the respective method and strategy. The canonical strategy produced no significant results in the positive side of the map; the same for the split-tails strategy. The combined strategy resulted in tests considered significant with both positive and negative test statistics. Since all p-values (for tests investigating positive and negative effects) are pooled in the correction, the thresholds are symmetrical. Here, fewer and smaller regions are identified as significant on the negative side (where there is a preponderance of signal) and some results on the positive side that were not seen with the canonical strategy appear significant. The two-tailed strategy produced broadly similar results as with the combined method. Finally, the split-tails strategy produced results that were visually generally similar as the canonical \textsc{fdr}. While the differences were minute in relation to \textsc{bh}, \textsc{bky} consistently produced slightly lower thresholds, leading to a slight increase in power. However, incorporating \textsc{bb} into the canonical and split-tails strategies resulted in thresholds remaining the same or further away from zero compared to their non-\textsc{bb} counterparts.


\section{Discussion}
\label{sec:discussion}

\subsection{Two-tailed tests}

Two-tailed tests trace their roots to the first decades of statistics as a scientific discipline. \citet[p.\ 47]{Fisher1925} noted that ``in some cases we wish to know the probability that the deviation, known to be positive, shall exceed an observed value, whereas in other cases the probability required is that a deviation, which is equally frequently positive and negative, shall exceed an observed value''. This allows assessment of extreme outcomes in any direction, not restricting hypotheses to a specific one. Similarly, in brain imaging, there are cases in which interest lies within only one of the tails, whereas in others, the interest lies within both tails of the distribution of test statistics. Consider, for example, an f\textsc{mri} study where the same information is presented via visual or auditory cues: some brain areas may be selectively modulated by vision and others by hearing; interesting comparisons involve both directions, i.e., higher involvement of some regions during visual than auditory, and vice-versa. Alternatively, consider a voxel-based morphometry study that compares gray matter volume between professional singers and professional pianists: again one might expect brain regions show differences in one group while other regions show the reverse pattern.

In effect, using two-tailed tests as default has at times been recommended for neuroimaging studies, in line with the fact that most studies investigate any differences between groups, regardless of direction. In most cases there is not enough {\it a priori} information to restrict comparisons to only one direction, hence \citet{Chen2019} argued that, ``it is difficult to imagine a scenario where [a] researcher would test all voxels simultaneously for just one directional change''. However, the researcher must be aware that the amount of errors is not guaranteed within a given direction if \textsc{fdr} is applied across all tests. Merely applying \textsc{fdr} methods as integrated into neuroimaging packages such as \textsc{afni}, \textsc{fsl}, FreeSurfer, or \textsc{spm} --- each referenced in thousands of peer-reviewed publications annually \citep{Poldrack2019} --- and then proceeding to directional inferences, does not control the error rate within each direction, as illustrated here using synthetic and real data.

\subsection{Control of the error rate}
\label{sec:discussion_control}

The analyses using synthetic data indicate that, while \textsc{fdr}-based adjustment (with either \textsc{bh} or \textsc{bky}) controls the error rate in the set of tests to which the procedure is applied, should the researcher decide to make inferences about direction of effects, extra care is necessary, otherwise such inferences can be affected by excess error rate. In cases where true signal is present in only one side, the error rate for the other side will be 100\% if any voxels are found significant there. Simple strategies such as using a two-tailed test followed by \textsc{fdr}, or combining the two sets of results with the directional hypotheses into one single \textsc{fdr} correction, do not guarantee control over the error rate for either direction, even though these control the error rate globally, i.e., when all tests are considered. Such global tests, however, provide no information on the direction of effects, i.e., no power to localize the direction of effects.

Alternative approaches were considered here. For example, running \textsc{fdr} separately for the two set of p-values with each of the directional hypotheses (the canonical strategy), or running \textsc{fdr} separately for the positive and negative test statistics (the split-tails strategy). The analyses demonstrate that these approaches are able to control the error rate in each direction. However, they come with an important penalty: should the null hypothesis be true everywhere, the error rate, which should be at the level $q$ under the complete null across both sides, is doubled, approaching $2q$ when both sides are considered. Without modification, should the researcher be interested in retaining the error controlled even in the complete null with these methods, then the test level should be halved, at the risk of loss of power. Trying to control \textsc{fdr} for all subsets simultaneously would transform it into an \textsc{fwer}-controlling procedure \citep{Finner2001}.

\subsection{Selective inference}

A more modest (but not less useful) goal, however, can be achieved: controlling the average false discovery error across multiple sets (families) of tests. By recasting the canonical and split-tails strategies as involving multiple such sets, the approach by \citet{Benjamini2014} (\textsc{bb}) comes to rescue: instead of blindly adapting the test level by dividing it by 2 to account for the doubled error rate under the complete null (observed in scenario \textsc{i} with canonical and split-tails), that adaptation is only applied to the extent that there is evidence allowing rejection of the null for a fraction of the sets. Moreover, the method accommodates not only two (for the two directions) but any number of ways in which the tests can be partitioned. Such partitioning can be based, for example, on regions of interest or parcellations of anatomical structures, as frequently used in brain imaging. And compared to their non-\textsc{bb} counterparts, the two-stage procedure does not sacrifice power, as demonstrated in Table~\ref{tab:synthetic_pwr} and Figure~\ref{fig:narps}.

\subsection{Localizing power}
\label{sec:discussion_localizing}

Even though the issues with lack of localizing power are a direct consequence of \textsc{fdr} not possessing the subsetting property described in Sec.~\ref{sec:theory_localizing}, routinely applied strategies for data analysis and visualization may obscure this fact. For example, the use of symmetrical thresholds is common in imaging visualization software: the user loads a statistical map and applies a threshold $t$; values above $+t$ are shown in a hot color scheme (i.e., red to yellow), while values below $-t$ are shown in a cold color scheme (i.e., blue to cyan); for two-tailed tests, this seems appropriate, clear and rigorous. Yet, the amount of errors in any particular direction is not controlled with \textsc{fdr}, only the global amount of errors, something that is not always considered by researchers, or they may (incorrectly) assume that the \textsc{fdr} is controlled in the positive and negative sides of the map. Another situation is when drawing conclusions about the amount of errors within a region after controlling globally; it might seem that controlling the error rate for the whole brain is the most rigorous, such that inferences within sub-regions would be conservative, given that a larger number of tests is taken into account. Yet, the amount of errors within any particular region is not controlled, only the global amount of errors. With currently used methods, any attempt to isolate a particular direction or a particular subset of tests will fail to guarantee the desired proportion of false discoveries within that subset.

Direction and localization errors can be avoided by refraining altogether from making localized or directional statements encompassing subsets of the tests, making only such statements when they refer to the entire set subjected to \textsc{fdr}. Or they can be avoided rigorously with the use of selective inference, i.e., with the \textsc{bb} procedure. For directional inferences, those that we named canonical + \textsc{bb} and split-tails + \textsc{bb} can be used.

\subsection{Real data}

While simulations using synthetic data allow quantification of the observed proportion of false discoveries and how it diverges from the nominal, expected rate, those using real data allow for the visualization of the different patterns that may be observed in typical studies, depending on the \textsc{fdr} method and correction strategy adopted. With real data, the canonical and split-tails led to broadly similar statistically significant spatial maps; likewise, the combined and two-tailed strategies were broadly similar one to another. Two-tailed tests can be seen as a particular case of multiple testing correction, since they consider both directions of an effect. Hence, it is not surprising that the two-tailed strategy produces nearly identical results as the combined strategy, which is similar to correction across contrasts \citep{Alberton2020}.

It should be understood, however, that areas of positive signal that appeared as significant with the two-tailed and combined strategies in the \textsc{narps} dataset example may have emerged simply because there is a preponderance of signal in the negative side of the map; evidence for that comes from the results using synthetic data. This is because procedures that control \textsc{fdr} are adaptive, that is, they start with a stringent control for the smallest p-value, and proceed in steps, each contingent on the findings of the previous one, progressively loosening stringency and continuing as long as each subsequent test continue to be declared significant \citep{Benjamini2000, Nichols2007}. With two-tailed and combined strategies, the large amount of tests with strong signal in one side of the map (negative in our example) allows more opportunities for tests in the opposite side (positive) to be declared significant. The canonical and split-tails strategies eliminate the opportunity for eventual strong or widespread signals present in one side impact the other, as \textsc{fdr} is applied separately to each side. When used with the \textsc{bb} procedure, significance $t$ thresholds are further increased, mitigating the risk of excess false positives that would occur under the complete null.

\subsection{More power with \textsc{bky}}
\label{sec:discussion_bky}

An important yet little appreciated fact is that the \textsc{bh} procedure --- the most commonly \textsc{fdr} method used --- does not control the amount of false discoveries at the exact level $q$ desired by the researcher, but at a lower level $qV_0/V$ that is determined by the (unknown) proportion of tests in which the null hypothesis is true among the total number of tests (Section~\ref{sec:bh}). Hence, the procedure is, in fact, conservative in relation to the level desired by the researcher. Despite this, it remains more powerful than procedures that control \textsc{fwer} such as \v{S}id\'ak or Bonferroni that use, for all tests, a fixed significance threshold that determines whether a p-value is significant; unless $V_0 = V$ (complete null), $\textsc{fwer}=1$ for any \textsc{fdr}-controlling procedure. Whereas \textsc{bh} is adaptive in the sense that the threshold varies according to the sorted p-value at position $i$, with a fixed slope given by $q/V$, the \textsc{bky} procedure is even more adaptive than \textsc{bh} in that the slope itself varies according to the sorted p-value at position $i$ as $q/(V+1-i(1-q))$ (Figure~\ref{fig:thresholds}). The quantity $q/(V+1-i(1-q))$ is only smaller than $q/V$ if $i<1/(1-q)$. Thus, for typical $q$, usually well below 0.5, \textsc{bky} is more powerful than \textsc{bh} as long as smallest ($i=1$) test is rejected. As far as power considerations are concerned, the \textsc{bky} procedure is therefore preferable over the \textsc{bh} procedure.

This extra power comes with a computational price: \textsc{bh} runs in $\mathcal{O}(V)$ time, whereas \textsc{bky} runs in $\mathcal{O}(V^2)$ time. This means that, while the time it takes to run \textsc{fdr} grows linearly with the number of tests to be corrected, for \textsc{bky} it grows quadratically. Another point worth mentioning is that \textsc{bky}-adjusted p-values can easily be above 1 (Figure~\ref{fig:adjusted}). This is not a unique feature of \textsc{bky}: other multiple testing procedures can also lead to adjusted p-values being higher than 1, the simplest being Bonferroni, that leads to adjusted p-values that are equal to the uncorrected p-values multiplied by the number of tests $V$. This does not have substantial implications for hypotheses testing, and can be solved via truncating adjusted values at 1.

The \textsc{bky} procedure was developed as an improvement over \textsc{bh}, and thus it would seem that the same considerations regarding dependencies between tests would hold. However, results have only been proven for the independent case \citep{Benjamini2006}, even if it tends to perform well in simulations under certain forms of dependence \citep[e.g.,][]{Stevens2017}. Scaling the thresholds of \textsc{bky} by $1/c(V)$ to accommodate arbitrary dependencies, as in \textsc{by}, while compelling, has not been proven to control the \textsc{fdr} either (nor that it does not).

\subsection{Dependencies among tests}

While only one case of dependencies among tests was considered, which resulted in adequate control over the error rate, existing results in the literature \citep{Genovese2002, Logan2004, Reiner-Benaim2007} demonstrate that, for typical cases, the modification proposed by \citet{Benjamini2001} for a more robust control over the false discovery proportion (the \textsc{by} procedure) is not necessary, and that even in the presence of negative correlations, \textsc{bh}, being more substantially more powerful, is preferable over \textsc{by}. That extends to \textsc{bky} as well.

The \textsc{bb} procedure is valid under two dependence cases: (\textit{a}) when sets are independent one from another (with any dependence structure within set), provided that the procedure in the first stage is local within that set, and (\textit{b}) when all p-values are \textsc{prds} on the subset of p-values corresponding to the true null hypotheses \citep[Theorems 1 and 3]{Benjamini2014}. Nonetheless, for the combined + \textsc{bb} method, which clearly violates \textsc{prds} (since $1-p_i$ is fully and negatively dependent of $p_i$), the procedure still performed well. The reason is that, although independence among sets is violated, that violation affects only p-values that do not contribute substantially to the selection of sets: for small $p_i$ (approaching zero), the corresponding $1-p_i$ is too far from being significant to affect the outcomes from the selection rule used in the first stage; for large $p_i$ (around 0.5), neither $p_i$ or $1-p_i$ is close enough to typical test levels (usually around 0.05) to cause an impact, and thus the selection procedure in stage 1 remains local.

\subsection{Visualization of results}
\label{sec:discussion_visualization}

The approaches that effectively control the false discovery rate in each one of the two sides of the map, or more generally in any partitioning of the map are straightforward for software developers to implement. However, their use in practice requires that the user is able to display the resulting statistical map with thresholds that are not constrained to be symmetric across both sides of the map. That is, the ideal imaging tool would accommodate separate thresholds for positive and for the negative sides of the map containg the test statistics, such that these can be set separately after running \textsc{fdr}. This also applies to approaches that do not use abrupt thresholds, but instead a transparency gradient, allowing subthreshold evidence be viewed informatively while highlighting suprathreshold results \citep{Allen2012, Taylor2025}.

Currently, very few software packages allow the user to easily set asymmetric thresholds for positive and negative sides of the map. Having this feature available across different imaging software would more easily allow for directional investigations with \textsc{fdr} and visualization of the relevant results. In the absence of software with such features, workarounds can be considered. For example, a map of uncorrected p-values can be thresholded at the \textsc{fdr}-controlling level, or a map of adjusted p-values can be thresholded at the desired test level, then displayed. Alternatively, if the distribution of the test statistic is known, the threshold p-value can be converted to a threshold statistic (using the cumulative distribution function, parametric or non-parametric), which can be applied to partition statistical map into significant and non-significant regions. A thresholded map (of statistics or of p-values) can be binarized and used as a mask to display or select regions for further analyses, possibly including sub-threshold results with transparency; the same workaround can be used for multiple regions, each of them subjected to correction separately, possibly with the test level further scaled down by the number of such regions should the researcher wish to control the error rate under the complete null.

\subsection{Recommendations}
\label{sec:discussion_recommendations}

The average error rate across sets of tests can be controlled successfully with the split-tails + \textsc{bb} strategy, while preventing direction errors. More generally, the main recommendation is that whenever there is interest in making statements about parts of a statistical map, the researcher should split it into disjoint sets and proceed with \textsc{bb} to ensure control over the average false discovery rate. A second recommendation is to use the \textsc{bky} procedure as routine, since it offers greater power than the \textsc{bh} procedure. A third recommendation concerns features in visualization software: to view results at the significance level that controls the error rate in each side (positive, negative) of the statistical map, software should ideally allow for thresholds to be asymmetric; without such feature, users need to manually mask images to view each side separately.


\section{Conclusion}

While \textsc{fdr} procedures provide more power than \textsc{fwer} methods if many tests are expected to be significant and the user is willing to accept some false positives, caution is needed when making directional inferences after using two-tailed tests. Using selective inference (\textsc{bb}), in which tests are split into sets, each subjected to local control and then all tests are corrected at an adjusted level, controls the expected average \textsc{fdr}, eliminates excess error rate, and generalizes to any form of localized inference. Using \textsc{bky} in lieu of \textsc{bh} as a general method to control the \textsc{fdr} yields greater power.

{\footnotesize
\section*{Source code}
\noindent
Code used for assessment of the various strategies and methods is available at:\\
\href{https://github.com/andersonwinkler/fdrfun}{https://github.com/andersonwinkler/fdrfun}.\\
Implementation of the methods for applied use is available in \textsc{palm} (Permutation Analysis of Linear Models):\\
\href{https://github.com/andersonwinkler/PALM}{https://github.com/andersonwinkler/PALM}\\
\par}

{\footnotesize
\section*{Author contributions}
\noindent
All authors contributed to all aspects of this research.
\par}

{\footnotesize
\section*{Competing interests}
\noindent
None.
\par}

{\footnotesize
\section*{Acknowledgements}
\noindent
Support was provided by the National Institutes of Health.
A.M.W.: R01-MH139547 and U54-HG013247; 
P.A.T.: ZIC-MH002888; 
T.E.N.: R01-EB026859, U24-DA041123, and U19-AG073585; 
C.R.: RF1-MH133701 and P50-DC014664. The contributions of the NIH author was made as part of his official duties as NIH federal employee, are in compliance with agency policy requirements, and are considered Works of the United States Government; however, the findings and conclusions presented in this paper are those of the authors and do not necessarily reflect the views of the NIH or the U.S. Department of Health and Human Services. The authors are grateful to the anonymous reviewers who provided helpful feedback.
\par}

{\footnotesize
\setlength{\bibsep}{0.0pt}
\begin{spacing}{0.9}
\bibliographystyle{elsarticle-harv}
\bibliography{refs.bib}
\end{spacing}}

\end{document}